# A Polarimetry-based Field-deployable Non-interruptive Mirror Soiling Detection Method

Mo Tian[1], Md Zubair Ebne Rafique[1], Kolappan Chidambaranathan[1], Randy Brost[2], Daniel Small[2], David Novick[2], Julius Yellowhair[3], Yu Yao[1,*]

[1] Electrical, Computer and Energy Engineering, Arizona State University. Tempe, AZ

2 Sandia National Laboratories. Albuquerque, NM

3 Gryphon Technologies. Albuquerque, NM

* Corresponding Author. Email address: yuyao@asu.edu

## Abstract

The soiling level of heliostat mirrors in Concentrated Solar Power (CSP) fields is one of the key factors that significantly influences optical efficiency. State-of-the-art methods of monitoring heliostats soiling levels still face various challenges, including slow speed, labor-intensive operations, resolution and accuracy constraints or interruptions to solar field operations. We present a rapid, cost-effective, and non-intrusive method for mirror soiling detection based on polarimetric imaging, referred to as Polarimetric Imaging-based Mirror Soiling (PIMS). The compact PIMS device is designed for integration with unmanned aerial vehicles (UAVs), enabling rapid, large-area assessments of heliostat mirrors for efficient soiling detection. Our method utilizes the correlation between the Degree of Linear Polarization (DoLP) and surface soiling level based on Mie scattering theory and Monte Carlo simulations. Field deployment of the PIMS method requires minimal device installation, and its UAV-based operation allows for soiling detection without interrupting plant activities. The PIMS method holds the potential for mirror soiling detection across various concentrated solar power (CSP) plants and can be further adapted for other types of solar fields, such as parabolic trough systems.

## 1. Introduction

Monitoring and improving optical efficiency are significant considerations in the operation of CSP fields. One of the most essential factors that influence optical efficiency is the soiling level of heliostat mirrors [1, 2]. The soiling of heliostat mirrors decreases the specular reflection of the mirror, and thus results in significant drop in optical efficiency [3, 4]. From various field studies [2, 5-11], the heliostat mirror soiling levels can change rapidly due to environmental factors and weather events, such as snow, rain, sandstorms, wind, dust, etc. Reflectance can change quickly due to these events in a single day, and long-term uncleaned mirrors experience a reflectance loss of approximately 10-15% as reported in [12]. Since it is usually difficult to predict the mirror soiling patterns at different times of a year, routine inspection and cleaning of the soiled mirror is crucial to the operation of the CSP field. Currently the inspection of heliostat mirror soiling in CSP plants relies on specular reflectometer. However, it does not provide the necessary combination of accuracy and speed. To maximize efficiency of monitoring and maintenance of heliostats in CSP fields, a desirable mirror soiling detection method for CSP field should be fast, low cost, not labor-intensive and not interrupt daily operation. So far, a few methods have been reported in literature for CSP field mirror soiling monitoring, such as irradiance measurement [13, 14] and machine vision [4, 15, 16]. However, each method still faces various limitations in the implementation during field operation. A summary of their strengths and limitations can be found in Table 1.

| Method | Strengths | Limitation |
| --- | --- | --- |
| TraCS: pyrheliometer-based system comparing direct and reflected Direct Normal Irradiance (DNI) [13] | High temporal resolution; uses real-time solar tracking to compare direct and reflected irradiance. | Only valid at a fixed sample location; does not follow the heliostat tracking path and cannot assess operational heliostats. |
| TraCS 2.0: motorized rotating mirror for DNI-based soiling detection [14] | Sensor-specific system; enables continuous DNI-based reflectance monitoring. | Limited to fixed sample mirrors; results may vary due to operator handling and environmental conditions. |
| AVUS: automated LED-based reflectance measurement of fixed mirror samples [4] | Precise, automated, low maintenance. | Limited to fixed sample mirrors; not representative of spatial variability across full heliostat field. |
| Drone imaging + sky modeling for reflectance estimation via RGB analysis [15] | Large-area scan, fast, detects soiling and corrosion. | Manual calibration; sensitivity to image quality and lighting |
| Dust-InSMS: CNN-based single RGB image analysis with GPS tagging [16] | High accuracy, real-time use. | Requires robust training; affected by environmental conditions; requires device installation on heliostats |

**Table 1 Summary of heliostat mirror soiling detection methods**

In this paper, we present a mirror soiling detection method to estimate the soiling level and corresponding reflectance of heliostat mirrors using polarimetric imaging. In this Polarimetric Imaging-based Mirror Soiling (PIMS) detection method, we have established an optical model to correlate polarization images of soiled mirrors to the mirror surface soiling levels and the corresponding relative reflectance (R). From an optical point of view, the major difference between a soiled mirror surface and a clean mirror surface is that the soiled particles have strong scattering effect on the incident light and thus change not only the intensity but also the polarization states of reflected light. Therefore, the relationship between the Degree of Linear Polarization (DoLP) of the captured images and relative reflectance of the soiled mirrors can be modeled using Mie scattering model combined with Monte Carlo simulation. The PIMS method was validated using a portable setup on samples prepared with different soil types, achieving a relative reflectance error between 1.41% and 2.77%. This error is defined as the percentage difference between reflectance predicted by PIMS and that measured using a custom laboratory-based specular reflectance system. As shown in Supplementary Fig. S5a, the setup uses a collimated white light source (400–700 nm), a pair of gold mirrors for beam steering, and an aperture-limited detection geometry with a power meter positioned at the specular angle. The clean region of the mirror is first measured and used as a baseline. The sample is then rotated to align different soiled regions with the beam spot, and the relative reflectance is calculated as the ratio of power meter readings (intensity) from soiled regions to the clean region. The system achieves <1% measurement uncertainty based on power meter specifications and repeatability under controlled laboratory conditions. We integrated a portable polarimetric imaging system on a UAV to take multiple single-shot images while the UAV flies over the heliostat field [17, 18]. In this way, the soiling level of different mirror facets captured in the image can be predicted directly without measuring each point with a reflectometer or installing complex devices on the heliostat field to monitor the soiling status, making the detection process fast, cost- efficient and labor-efficient.

Furthermore, PIMS uses skylight and sunlight as light source, minimizing installation requirements and avoiding interruption to field operation during the soiling detection. Our previous work [18] introduced the basic concepts of the PIMS method. In this paper, we extend the model to account for different soil types, enabling adaptation to different fields. We also present full DoLP simulations for slanted and heliostat mirrors, along with measurement results from soiled mirror samples, slanted mirrors, and heliostat field tests.

We demonstrated the effectiveness of this method in estimating mirror soiling levels using polarization images during field tests conducted at the Sandia National Solar Thermal Test Facility (NSTTF). Compared to most state-of-the-art reflectance measurement methods for mirror soiling assessment in CSP fields, the PIMS method offers several advantages, including high throughput measurement (multiple heliostats at a single snapshot), minimal interference with field operations, simple and high-speed data processing, and minimal additional equipment installation. While conditions vary across different solar fields, the PIMS method is adaptable to various setups, including vehicle integration and handheld configurations.

## 2. Concept and Model Formulation

In addition to intensity and wavelength, light also has the crucial property of polarization. The polarization states of light can be fully described using Stokes parameters, $(s_0, s_1, s_2, s_3)^T$. A partially polarized light can be decomposed into an unpolarized component and polarized component, as described in Equation (1).

$$S = \begin{pmatrix} s_0 \\ s_1 \\ s_2 \\ s_3 \end{pmatrix} = (1 - \mathcal{P}) \begin{pmatrix} s_0 \\ 0 \\ 0 \\ 0 \end{pmatrix} + \mathcal{P} \begin{pmatrix} s_0 \\ s_1 \\ s_2 \\ s_3 \end{pmatrix} \qquad (1)$$

$$\mathcal{P} = \frac{\sqrt{s_1^2 + s_2^2 + s_3^2}}{s_0} \qquad (2)$$

Here, the degree of polarization $\mathcal{P}$ can be calculated using Equation (2). The first term in Equation (1) represents the unpolarized component of light, which is only related to intensity, while the second term accounts for the polarized components of the light. If there is no circular polarization component, $s_3 = 0$. For completely unpolarized light, the degree of polarization equals zero. When considering only the linear polarization components, the degree of linear polarization (DoLP) is used to describe the state of polarization, as in Equation (3). DoLP has a value range from 0 to 1. When the light is completely linearly polarized, $DoLP = 1$.

$$DoLP = \frac{\sqrt{s_1^2 + s_2^2}}{s_0} \qquad (3)$$

## 2.1. Skylight Polarization

The sunlight, after traveling through the atmosphere, forms the skylight that covers every direction of the sky dome as a result of Rayleigh scattering. As observers standing on the ground, we see skylight from all directions on the sky dome, as shown in Fig.1a. The pattern of skylight DoLP is dependent on longitude, latitude, date and time [18]. Date and time influence the Sun position on the sky dome, described in the coordinates defined in Supplementary Fig.S1. In accordance with our field test facility NSTTF's coordinates, the azimuthal angle starts at North as 0 and increases clockwise while the zenith angle is 0 at zenith and 90 when the vector is parallel to the ground. Thus, given a specific Sun position and skylight incident angle, the incident light's polarization states can be calculated from Rayleigh scattering. For example, when Sun is at zenith $\theta_{sun} = 30°$ and azimuth $\phi_{sun} = 90°$, the skylight DoLP has the pattern as shown in Fig.1c. $\theta_{sun}$ The direct sunlight incidence direction has the lowest DoLP as sunlight is considered unpolarized.

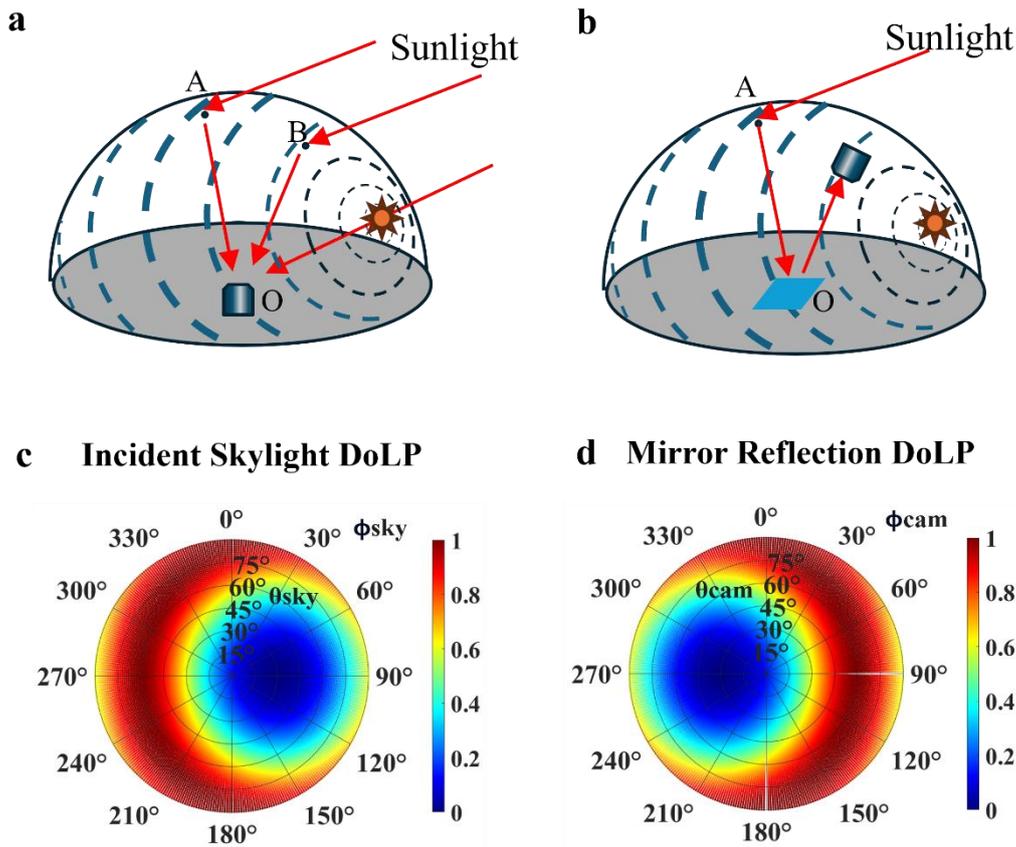

**Fig.1 Skylight DoLP due to Rayleigh scattering. a,** Diagram illustrating the formation of skylight through the scattering of sunlight in the atmosphere. **b,** Diagram illustrating the reflection of skylight from a horizontally oriented clean mirror surface, with its normal pointing vertically upward. **c,** Incident skylight DoLP given $\theta_{sun} = 30°$ and $\phi_{sun} = 90°$. The coordinates represent the incident skylight angles, $\theta_{sky}$ and $\phi_{sky}$. **d,** Simulated DoLP pattern of skylight reflected from a horizontally positioned mirror. The coordinates thus represent the camera position, $\theta_{cam}$ and $\phi_{cam}$.

## 2.2. Polarization Image of Clear Sky in Clean Mirror Reflection

Based on the incident skylight DoLP pattern, if a reflective mirror is positioned horizontally with its surface normal aligned with the zenith direction, one can calculate the reflected light's polarization states at different angles using a Reflection Mueller Matrix [19], as shown in Fig.1d. Now the coordinates describe the camera's zenith and azimuthal angle $\theta_{cam}$ and $\phi_{cam}$, with the center of the leveled mirror as the origin.

The derivation of the Reflection Mueller Matrix begins with the definition of the Stokes parameters and the Jones-to-Stokes transformation. Given the parallel and perpendicular components of the electric field ($E_x$ and $E_y$), the Stokes parameters can be defined as follows:

$$S = \begin{pmatrix} S_0 \\ S_1 \\ S_2 \\ S_3 \end{pmatrix} = \begin{pmatrix} |E_x|^2 + |E_y|^2 \\ |E_x|^2 - |E_y|^2 \\ 2\,\text{Re}\{E_x E_y^*\} \\ -2\,\text{Im}\{E_x E_y^*\} \end{pmatrix} \quad (4)$$

where $E_x$ and $E_y$ represent the complex amplitudes of the electric field in two orthogonal polarization directions, and the superscript (*) denotes the complex conjugate. From Fresnel's equations [19], one can construct a Jones reflection matrix for fully polarized incident light as:

$$J_{ab,R} = \begin{pmatrix} R_l & 0 \\ 0 & R_r \end{pmatrix} \quad (5)$$

$$R_l = \frac{n_b \cos\theta_a - n_a \cos\theta_b}{n_b \cos\theta_a + n_a \cos\theta_b} \quad (6)$$

$$R_r = \frac{n_a \cos\theta_a - n_b \cos\theta_b}{n_a \cos\theta_a + n_b \cos\theta_b} \quad (7)$$

$R_l$ and $R_r$ represent the parallel and perpendicular reflection coefficients of the Fresnel Equations at the interface between the two media, respectively. $n_a$ and $n_b$ are the refractive indices of the two media. $\theta_a$ and $\theta_b$ are the incident and reflection angles. In the specular reflection case, the incident and reflection angles are the same, i.e., $\theta_a = \theta_b$. However, the Jones formalism applies only to fully polarized light. To handle partially polarized and unpolarized light, we use Mueller matrix formalism.

$$M = A \cdot (J \otimes J^*) \cdot A^{-1} \quad (8)$$

$$A = \begin{pmatrix} 1 & 0 & 0 & 1 \\ 1 & 0 & 0 & -1 \\ 0 & 1 & 1 & 0 \\ 0 & i & -i & 0 \end{pmatrix} \quad (9)$$

where $\otimes$ denotes the Kronecker product and A is a transformation matrix that converts the Jones vector representation into the Stokes vector representation. Thus, we have

$$S_R = M_{ab,R} \cdot S_I \quad (10)$$

$$M_{ab,R} = \begin{pmatrix} \frac{1}{2}(R_l R_l^* + R_r R_r^*) & \frac{1}{2}(R_l R_l^* - R_r R_r^*) & 0 & 0 \\ \frac{1}{2}(R_l R_l^* - R_r R_r^*) & \frac{1}{2}(R_l R_l^* + R_r R_r^*) & 0 & 0 \\ 0 & 0 & Re\{R_l R_r^*\} & Im\{R_l R_r^*\} \\ 0 & 0 & -Im\{R_l R_r^*\} & Re\{R_l R_r^*\} \end{pmatrix} \quad (11)$$

Here, $M_{ab,R}$ denotes the Reflection Mueller Matrix from medium a to medium b. $S_R$ and $S_I$ represent the Stokes parameters of reflected and incident light.

### 2.3. *Polarization Image of Clear Sky in Soiled Mirror Reflection*

To predict reflectance of soiled surface using polarization images, we developed an optical model to simulate the process of light reflection and scattering. These two processes have different impacts on the change of polarization states. If the incident light is highly linearly polarized (i.e., exhibits a high Degree of Linear Polarization, DoLP), as illustrated in Fig. 2a, the specular reflection will also retain a high DoLP. This preservation of polarization can be quantitatively described using the Reflection Mueller Matrix, as defined in Equation (11). In contrast, when the incident light interacts with soil particles on the mirror surface, the scattering process leads to significantly reduced DoLP in the diffusely reflected light. The soiled particles size on heliostat mirrors of a solar field are typically in the range of sub-micrometer to 1 millimeter [20], which are generally similar or larger than the wavelength of visible light. From Mie scattering theory [21-23], we can calculate the change of polarization states and scattering angle during a single scattering event. Using Monte-Carlo simulation, we describe the scattering angle profile as a probability distribution in 3-dimensional space.

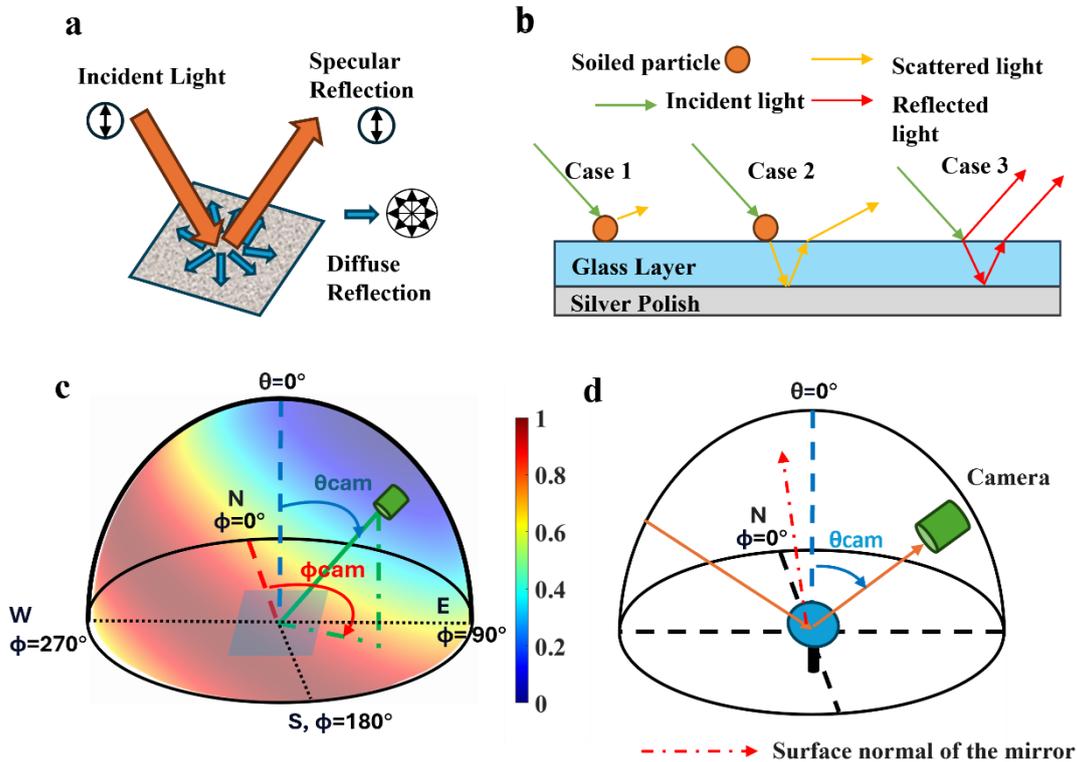

**Fig.2 PIMS Simulation Model to Calculate Relative Reflectance Using DoLP. a,** Illustration of the interaction between a highly linearly polarized incident light and a soiled reflective surface. The orange arrows represent incident and reflected beams, both with a high Degree of Linear Polarization (DoLP), as indicated by the uniform vertical polarization symbol. In contrast, the diffuse reflection represented by blue arrows with varied directions results from scattering by surface soiling and exhibits low DoLP due to depolarization. **b,** Different scenarios of incident light encounter the mirror surface. From left to right, case 1: scattered light goes above the mirror surface; case 2: scattered light transmits into the glass layer and gets reflected by the silver polish; case 3: light incident on the clean part of the mirror surface and gets reflected. **c,** the defined coordinate systems for camera, defined by zenith angle $\theta$ and azimuthal angle $\phi$. The same coordinate system is used in all the other simulations and tests. **d,** Diagram showing the simulation of the Skylight pattern in spherical coordinates as a dome.

As previously defined in Section 2.1 and Supplementary Fig. S1, we adopt a global coordinate system at the Sandia NSTTF, with azimuthal angle measured clockwise from geographic north (0°) and zenith angle defined from vertical (0° at zenith, 90° at the horizon). To model skylight polarization consistently across the heliostat field, we apply this global coordinate frame regardless of whether the origin is placed at the solar tower, a heliostat, or another reference point. The Sun's position on the celestial dome is determined by geographic coordinates (latitude and longitude) and time, and can be precisely calculated. Over short spatial separations within the field (e.g., a 0.1° angular change corresponds to ~11 km on the ground), the relative position of the Sun and the corresponding skylight DoLP pattern remain effectively constant. Therefore, minor translations in origin do not significantly affect skylight polarization modeling. This geometric invariance enables a consistent treatment of incident skylight across mirrors and allows all polarization calculations to be referenced to a unified coordinate system.

When an image of the soiled mirror was taken, the camera is collecting a near-specular reflection light within its lens acceptance angle at 9 degrees (or 157 mrads) [1, 3]. Thus, both specular reflection and part of the scattered light will be collected by the camera. Qualitatively, when the incident light has a high DoLP value, more soiled particle coverage on the mirror surface results in more scattering, and thus lower DoLP. As mentioned above, the incident light needs to be highly linearly polarized. To utilize this method on a CSP field with large area scanning, we can use the higher DoLP regions of the skylight as it naturally has a polarization pattern, as shown in Fig.2c. When the incident light arrives at the mirror surface, it gets reflected by the mirror surface or scattered by the soiled particles on the mirror, as shown in Fig.2b. Notice that direct sunlight is mainly unpolarized with very low DoLP, and the band pattern 180° away in azimuth angle from the Sun has a high DoLP. To get the best contrast, we placed our camera in the position and orientation such that the region of the sky seen reflected in the mirror has relatively predicted high DoLP, from the simulation of skylight. As time and location influence the sun position on the celestial dome, this DoLP simulation pattern changes for each measurement, and the suitable camera positions must be chosen differently according to each simulation. In PIMS model, the soiled mirror can be seen as a mirror surface with multiple spherical particles in different sizes deposited on it. In Mie theory, the scattering coefficient is only dependent on the scattering cross-section and the refractive index of the particle. We treat these particles as spherical particles with the same refractive index assuming they are the same kind of soil. Statistically, we are calculating the summation of the scattering events of each particle, thus the individual particle's shape and refractive index difference were neglected for simplification. Supplementary Fig.S2 shows the particle size distribution of different types of soil measured using SEM image processing [18], Dynamic Light Scattering (Malvern Zetasizer Nano) and Nanoparticle Tracking Analysis (Malvern Panalytical NanoSight NS300). Fig.2d illustrates the case when the measured mirror is slanted at an angle. If we define the surface normal vector of the mirror to be $\vec{N} = (\theta_{mirror}, \phi_{mirror})$, to get the desired skylight with high DoLP with certain incidence vector $\vec{Inc} = (\theta_{sky}, \phi_{sky})$, we need to place the camera at position:

$$\vec{Cam} = (\theta_{cam}, \phi_{cam}) = \vec{Inc} - 2(\vec{Inc} \cdot \vec{N})\vec{N} \qquad (12)$$

Here, we define the origin of the coordinate system as the geometric center of the slanted mirror surface.

We consider the light collected by the camera in two parts [18]. First, for the area of the mirror that is not covered by any soiled particles, we consider it directly reflects the skylight of the corresponding angles, and the Stokes Parameters change during this reflection process can be calculated using Muller Matrix as in Equation (11). In the third case shown in Fig.2b, when the mirror has more than one layer, it is also possible for the light to transmit through the interface between air and the glass layer, then get reflected by the silver polish. In addition to the Reflection Mueller Matrix, a Transmission Mueller Matrix [19] is then used to calculate the change of polarization states during this process, as described in Equations (13-17).

$$S_T = M_{ab,T} \cdot S_I \quad (13)$$

$$M_{ab,T} = f_T \begin{pmatrix} \frac{1}{2}(T_l T_l^* + T_r T_r^*) & \frac{1}{2}(T_l T_l^* - T_r T_r^*) & 0 & 0 \\ \frac{1}{2}(T_l R_l^* - R_r R_r^*) & \frac{1}{2}(T_l T_l^* + T_r T_r^*) & 0 & 0 \\ 0 & 0 & Re\{T_l T_r^*\} & Im\{T_l T_r^*\} \\ 0 & 0 & -Im\{T_l T_r^*\} & Re\{T_l T_r^*\} \end{pmatrix} \quad (14)$$

$$f_T = n_{ba}^3 \left(\frac{\cos\theta_b}{\cos\theta_a}\right) \quad (15)$$

$$T_l = \frac{2n_a \cos\theta_a}{n_b \cos\theta_a + n_a \cos\theta_b} \quad (16)$$

$$T_r = \frac{2n_a \cos\theta_a}{n_a \cos\theta_a + n_b \cos\theta_b} \quad (17)$$

Similar to reflection Mueller Matrix, $M_{ab,T}$ denotes the Transmission Mueller Matrix from medium a to medium b. $S_T$ and $S_I$ represent the Stokes parameters of transmitted and incident light. $T_l$ and $T_r$ represent the parallel and perpendicular transmission coefficients of the Fresnel Equations at the interface between the two media, respectively. $n_a$ and $n_b$ are the refractive indices of the two media. $\theta_a$ and $\theta_b$ are the incident and reflection angles. In the specular reflection case, the incident and reflection angles are the same, i.e., $\theta_a = \theta_b$.

Second, for the area of the mirror that is covered by the soiled particles, the scattered sunlight is dominant as the sunlight intensity is much higher than the skylight [25]. The scattering of skylight is not considered in PIMS model, as it adds no significant change to the results but will be a large computational drag. In Supplementary Fig.S3, the simulation results of adding these skylight scattering components were demonstrated and there was no significant difference. For the scattering of sunlight, some of the sunlight will be backscattered into the viewing angle of the camera and thus is collected by the sensor, as in the first case in Fig.2b. Some will be forward scattered and reach the back of the silver coating, and then get reflected, as shown in the second case in Fig.2b. The sum of these two sources of collected light forms the camera image and we can then model the Degree of Linear Polarization as a function of soiling level [18] as expressed in Equation (18).

$$S_{total} = S_{sun,scat}\, A_P K_{sun} + S_{sky,refl}\, A_{NP} K_{sky} \qquad (18)$$

$S_{total}$ is the total Stokes Parameter. $S_{sun,scat}$ is the Sunlight's Stokes Parameter after scattering off the soiled particle and $S_{sky,refl}$ is the Skylight's Stokes Parameter after reflection off the clean mirror. Here, $S_{sun,scat}$ is calculated using Mie scattering theory and Monte Carlo simulation [18]. $S_{sky,refl}$ is calculated using Reflection Mueller Matrix as described in Equation (11). $A_P$ and $A_{NP}$ denote the Area Percentage of the sample covered by soil particles and not covered by soil particles, respectively, adding up to 100%. These values are derived from the previously measured particle size distribution. Specifically, we estimate the particle count density (in m⁻²), assume circular particle shapes based on measured sizes, and compute the total particle-covered area relative to the image area in a single frame. $K_{sun}$ and $K_{sky}$ denote the Sunlight and Skylight intensity ratio, adding up to 100%.

With the calculated Stokes Parameters collected by the camera (see Section 3.3 for explanation of the polarization image), we can then determine the relationship between the mirror reflectance and the DoLP of reflected light. In both simulation and measurement, the reflectance is considered at visible light wavelength range (400nm to 700nm). Several mentioned parameters such as the mirror structure and material, soiling particles size distribution and material can be fine-tuned according to the actual conditions of the testing facility or CSP field. The relative reflectance R is defined as the ratio between the clean region's total reflected intensity and soiled region's total reflected intensity taken in different images, as shown in Equation (19). Here, we define a near-specular reflection with an acceptance angle of less than 157 mrad for the lab-customized optical setup (corresponding to 9°; see Supplementary Fig. S5a) and less than 52 mrad for the Surface Optics 410-Solar reflectometer (corresponding to 3°; see Supplementary Fig. S5b). To maintain consistency with the polarization camera's operating range, the custom laboratory-based specular reflectance system was used with a light source and power meter covering the 400–700 nm wavelength range. In the field tests, we used the Surface Optics 410-Solar reflectometer, which measures reflectance over multiple discrete wavelength bands. For consistency with the laboratory measurements and polarization camera response, we selected and averaged only the three reflectometer bands: 400–540 nm, 480–600 nm, and 590–720 nm. Since for the ideally clean mirror, there are no soiled particles to cause scattering events, the total Stokes Parameters for the clean region's reflection are simplified to Equation (20). For mirror surface with higher soiling levels, $A_P$ is larger and $(S_{sun})_{scat}$ changes due to more scattering events in Equation (18), resulting in a lower R.

$$R = \frac{S_{0,total}^{soiled}}{S_{0,total}^{clean}} \quad (19)$$

$$S_{total}^{clean} = S_{sky,refl} K_{sky} \quad (20)$$

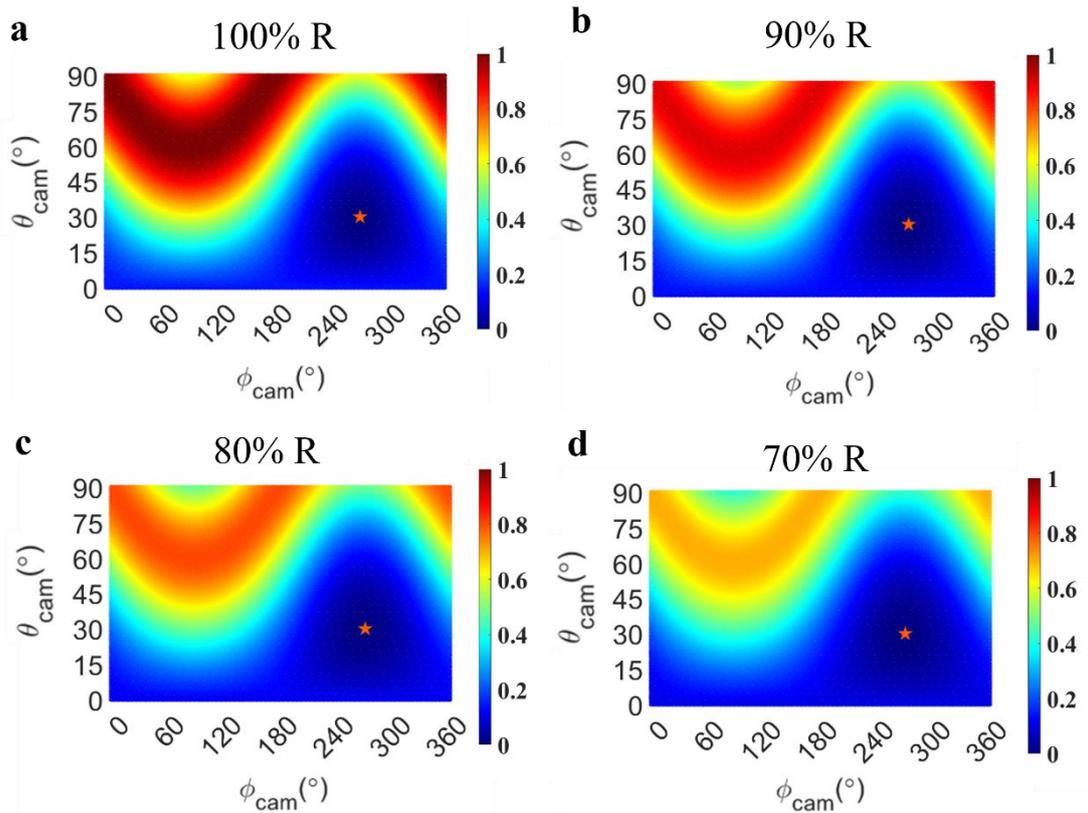

**Fig.3 Reflection DoLP simulation of different relative reflectance. a,** clean region, R=100%. **b-d,** R=90%, 80% and 70%**.** The stars indicate the camera position when looking directly at Sun reflection and getting lowest DoLP, at $\theta_{cam} = 30°$ and $\phi_{cam} = 270°$.

Figure 3 presents a DoLP reflection simulation based on the geometry defined in Fig. 2d. A clean mirror is modeled as a horizontally oriented planar surface with its surface normal aligned along the global +z axis. The camera is positioned at various azimuthal and zenith angles relative to the mirror's coordinate frame to simulate the reflected skylight arriving from different incident directions. Figure 3a-3d demonstrate how the reflection DoLP pattern from the mirror changes as the relative reflectance changes in PIMS simulation, corresponding to R=100%, 90%, 80% and 70%, respectively. Relative reflectance is defined as the same as Equation (19), with an acceptance angle of 10 degrees diagonal. In this simulation, we modeled the mirror as an optically smooth Ag film covered by glass (refractive index data [28]), refractive index of air as 1 and soiled particles as silica spheres (refractive index data [29]). The Sun position was set at $\theta_{sun} = 30°$ and $\phi_{sun} = 90°$. Now the mirror is modeled as a horizontally oriented planar surface, with its surface normal aligned along the +z axis in the mirror's local coordinate system, we have $\theta_{mirror} = 0°$. We define the geometric center of the mirror surface as the origin of the coordinate system, with the surface normal aligned along the +z axis and the mirror plane lying in the x–y plane. Thus, these simulation figures show the theoretical DoLP values that the camera is going to capture at different positions. On the reflected DoLP pattern the lowest DoLP was found at $\theta_{cam} = 30°$ and $\phi_{cam} = 270°$, corresponding to the specular reflection of sunlight, as indicated on the figures. These simulation results directly show that as the reflectance decreases the DoLP values decrease overall yet still have distinguishable patterns. These areas with high DoLP values are desired for applying this method in the measurement and field test.

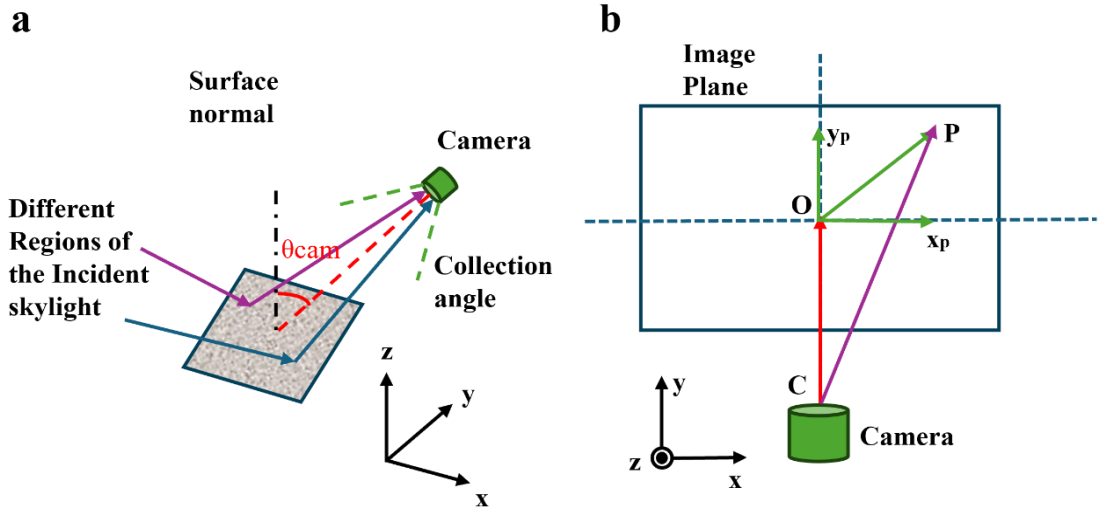

**Fig.4 Acceptance Angle Correction. a,** Global coordinate system showing the camera location and varying incident skylight directions across the field of view. Even within a single image, the polarization state of skylight can differ across the scene due to variations in the camera's acceptance angle. **b,** Local coordinate system showing the camera image plane. Each pixel point P can be mapped to a vector in global coordinates, enabling independent calculation of the corresponding incident skylight direction and polarization state.

When the center line-of-sight of the camera has the $\theta_{cam}$ and $\phi_{cam}$ reflected from the highest incidence DoLP, different sections of the image captured can still have different DoLP incidence due to the camera acceptance angle, as shown in Fig.4a. In this case, it is also useful to recalculate each region's incidence DoLP accordingly acquire more accurate results. This Acceptance Angle Correction (AAC) is applied in the later test results discussed in Section 5. To calculate the reflected vector CP in Fig.4b using global coordinates, we first convert the coordinates of OP from the local coordinate system defined in Fig. 4b to the global coordinate system shown in Fig. 4a, where the x-axis points east, the y-axis points north, and the z-axis points upward toward the zenith. If we set vector OP as $\vec{v}_{global}$ in global coordinates and $\vec{v}_{cam}$ in camera imaging plane coordinates, we have:

$$\vec{v}_{global} = R_{cam\_to\_global} \cdot \vec{v}_{cam} \quad (21)$$

$$R_{cam\_to\_global} = [\hat{x}_{cam}^{global} \quad \hat{y}_{cam}^{global} \quad \hat{z}_{cam}^{global}] \quad (22)$$

$$\hat{z}_{cam}^{global} = \frac{\overrightarrow{CO} - \overrightarrow{CAM}}{\|\overrightarrow{CO} - \overrightarrow{CAM}\|} \quad (23)$$

$$\hat{x}_{cam}^{global} = \frac{z' \times \hat{z}_{cam}^{global}}{\|z' \times \hat{z}_{cam}^{global}\|} \quad (24)$$

$$\hat{y}_{cam}^{global} = \hat{z}_{cam}^{global} \times \hat{x}_{cam}^{global} \quad (25)$$

$R_{cam\_to\_global}$ is the rotation matrix that converts imaging plane coordinates to global coordinates. $\hat{x}_{cam}^{global}$, $\hat{y}_{cam}^{global}$, and $\hat{z}_{cam}^{global}$ are the conversion equation for each axis. $\overrightarrow{CO}$ is the vector from camera to image plane center, as indicated in Fig.4b. $\overrightarrow{CAM}$ is camera's positional vector in global coordinates. $z'$ is the unit "up" vector in global coordinates. $z' = [0, 0, 1]$. Then, we can calculate the reflection vector at point P as,

$$\overrightarrow{CP} = \overrightarrow{CO} + \overrightarrow{OP} = \overrightarrow{CO} + \vec{v}_{global} \quad (26)$$

With known surface normal of the mirror, whether leveled or tilted at an angle, we can apply Equation (12) by switching the incident and reflection vectors to calculate the vector of incidence skylight and find its corresponding DoLP in the simulation. While we prefer using regions with the highest DoLP for optimal contrast, PIMS remains effective even when the incident light comes from areas with slightly lower DoLP. This principle applies not only to a single image with varying incident angles at different regions, as illustrated in Fig. 4a, but also when the entire image is captured under skylight with a generally lower DoLP. As illustrated In Supplementary Fig.S4, a measurement at different azimuthal angles and the same zenith angles were carried out. While at different angles the DoLP values are different for the same soiling region on the sample, the variation of reflected DoLP can be predicted by the simulation.

## 3. Detection Method and Setup

To validate the PIMS method prior to conducting field tests, we performed a series of measurements using soiled mirror samples on the top floor of Arizona State University's parking garage. The samples were prepared with various types of soil to assess the feasibility of deploying PIMS in fields with differing soil compositions. For each measurement, we prepared multiple samples using the same type of soil. Samples are round silver mirrors with different soiling levels in three soiled regions and one clean region. Based on the simulation of skylight DoLP pattern at different times and location, the angle ranges were selected to have the relatively high DoLP after reflecting from a soiled mirror. Next, the polarization camera (Allied Vision Mako G-508B POL) set on a tripod was placed to the desired azimuthal and zenith angle. After adjusting the focus of the lens and locating the mirror to the center of the image frame, we captured raw images and obtained DoLP images of mirror samples as described in Section 4.1. Among various images taken at different angles during the scanning process, we selected the image with the highest DoLP values in the clean region to obtain the data points for the DoLP in each region (see in Supplementary Fig.S4).

On the same day as the DoLP image acquisition, the relative reflectance of each mirror region was also measured using a custom laboratory-based specular reflectance system (see Supplementary Fig. S5a). The system operates over the 400–700 nm spectral range, with an acceptance angle of approximately 157 mrads and a fixed incidence angle matching the DoLP imaging geometry, as described in Section 3.2. The relative reflectance is defined as the ratio of reflected power from each soiled region to that of the clean reference region, as expressed in Equation (19).

### 3.1. Sample Preparation

The soiled mirror samples tested at ASU parking lot were prepared in the laboratory using different types of soil as discussed below with an enclosed chamber and a blow dryer. After the blow dryer blew the soil particles in the air above the mirror, the blow dryer was unplugged, and the clean mirror sample was placed in the chamber for different periods of time to prepare different levels of soiling. To ensure PIMS method can be applied to different fields, we collected different soil (of different refractive indices and particle size distributions) to prepare our samples. We prepared and measured four types of samples, named after the soil type as "Sahara White (SW)", "Sahara Red (SR)", "Sandia Soil (SAND)", "Sandia Heliostat Samples (HELIO)". The first three types of samples were prepared using a round mirror with a silver coating. However, the Sandia Heliostat Samples were prepared using small-sized, actual heliostat mirror segments obtained from Sandia NSTTF. These mirrors have different layers of coating and maintained their original soiling during shipment. For Mie scattering calculations, different types of soil were acquired and tested using Dynamic Light Scattering to get the particle size distributions. A later improvement of the sample preparation was introduced using a deposit chamber to mimic the natural soil deposition on solar field. This chamber first heats up the space with the soil particles and air, then cools down rapidly to form dew on the surface of the mirror. Next, the mirror is heated up to dry, and the soil contained in the dew is deposited on the surface of the mirror. With the help of this chamber, each region's soiling is more uniform and consistent, making it easier to acquire the DoLP data with small standard deviations.

### *3.2. Reflection Measurement*

We designed a custom optical setup to measure the specular reflectance of the mirror samples prepared at ASU (see Supplementary Fig. S5a). On an optical table, an unpolarized white light source with a spectral range of 400–700 nm was aligned through a set of apertures and a condenser lens to produce a collimated beam with a spot size of approximately 1 inch in diameter and a divergence angle of 0.5°. The beam was redirected by two gold-coated mirrors to maintain a fixed incidence angle matching the geometry used during DoLP image acquisition. The reflected beam was collected by a power meter (with a 0.9-inch diameter sensor, nearly matching the beam spot size), positioned at the specular reflection angle. The acceptance angle of this detection geometry was approximately 9° (157 mrad), defined by the optical alignment and aperture stop. During measurement, ambient lighting in the laboratory was turned off to eliminate background interference, and power readings were recorded only after stabilization. For each sample, the clean region was measured first to establish a baseline reflectance. The sample was then rotated to position different soiled regions into the measurement spot without altering the optical alignment. The relative reflectance was computed as the ratio of reflected power from each soiled region to the clean region under the same measurement conditions. Each measurement was repeated three times and averaged to reduce variability. This setup enables precise and consistent reflectance measurements for validating the PIMS method under controlled optical conditions.

In the field test, instead of using the lab-customized optical setup, a Surface Optics 410-Solar reflectometer was first calibrated with its reference coupon and a small heliostat mirror sample. Then we held it tightly against the heliostat mirror and repetitively took measurements 3 to 5 times and used the average as the data. Even so, because the reflectometer only relies on human operators to hold it stably, there were inevitably several measured spots ending with larger standard deviations or totally invalid results, for example, with specular reflection reported a negative value. Data points with standard deviation larger than 5% were all omitted from the data processing as there were errors introduced by operation. We conclude that our use of a reflectometer in the field to get the reflectance data was not efficient nor consistently reliable. To ensure consistency with our simulation model and polarization images, we use the Surface Optics 410-Solar reflectometer data by averaging three measured spectral bands: 400–540 nm, 480–600 nm, and 590–720 nm.

### 3.3. Polarization Image Processing

The raw image taken with the polarization camera (LUCID Triton) contains the intensity information of the four linear gratings at 0 degree, 45 degrees, 90 degrees and 135 degrees [26]. Stokes parameters are calculated below in Equation (27). DoLP is then calculated using Equation (3).

$$\begin{bmatrix} S_0 \\ S_1 \\ S_2 \end{bmatrix} = \begin{bmatrix} \frac{(I_0 + I_{90}) + (I_{45} + I_{135})}{2} \\ I_0 - I_{90} \\ I_{45} - I_{135} \end{bmatrix} \quad (27)$$

After processing the DoLP images, it is important to match the region where the specular reflection measurement was taken. On the small mirror samples, it can be measured easily with a ruler. However, on the heliostats, this will be rather difficult as the Surface Optics 410-Solar Reflectometer spot size is less than 1 inch, and the heliostats mirrors are 48 inches by 48 inches. During our measurement, we left markers and recorded the positions on the heliostat mirrors to retrieve the accurate position for acquiring the DoLP data. These were used in our measurement to evaluate the accuracy of the method.

## 4. Results of Outdoor Measurements
### 4.1. Measurements on Leveled Mirror Samples

Figure 5a shows a DoLP simulation of reflection from a mirror modeled as a horizontally oriented planar surface, with its surface normal aligned along the +z axis in the mirror's local coordinate system. $\theta_{cam}$ and $\phi_{cam}$ are the zenith and azimuthal angles of the camera, which provides us with guidance of where to position the camera for highest DoLP. Fig.5b is an image of the setup in measurement. For the four different sections with different soiling levels, we expect the DoLP values to be also different, as shown in Fig.5c. In the Monte Carlo simulation, different soiling levels for regions A–D were modeled by varying the coverage of soil particles on the reflective surface. As the coverage increased, both the Degree of Linear Polarization (DoLP) and the reflectance decreased, with reflectance values ranging from 70% to 100%. This simulation provides predictive insight into how DoLP varies with soiling, which we later validate through field measurements shown in Fig. 6.

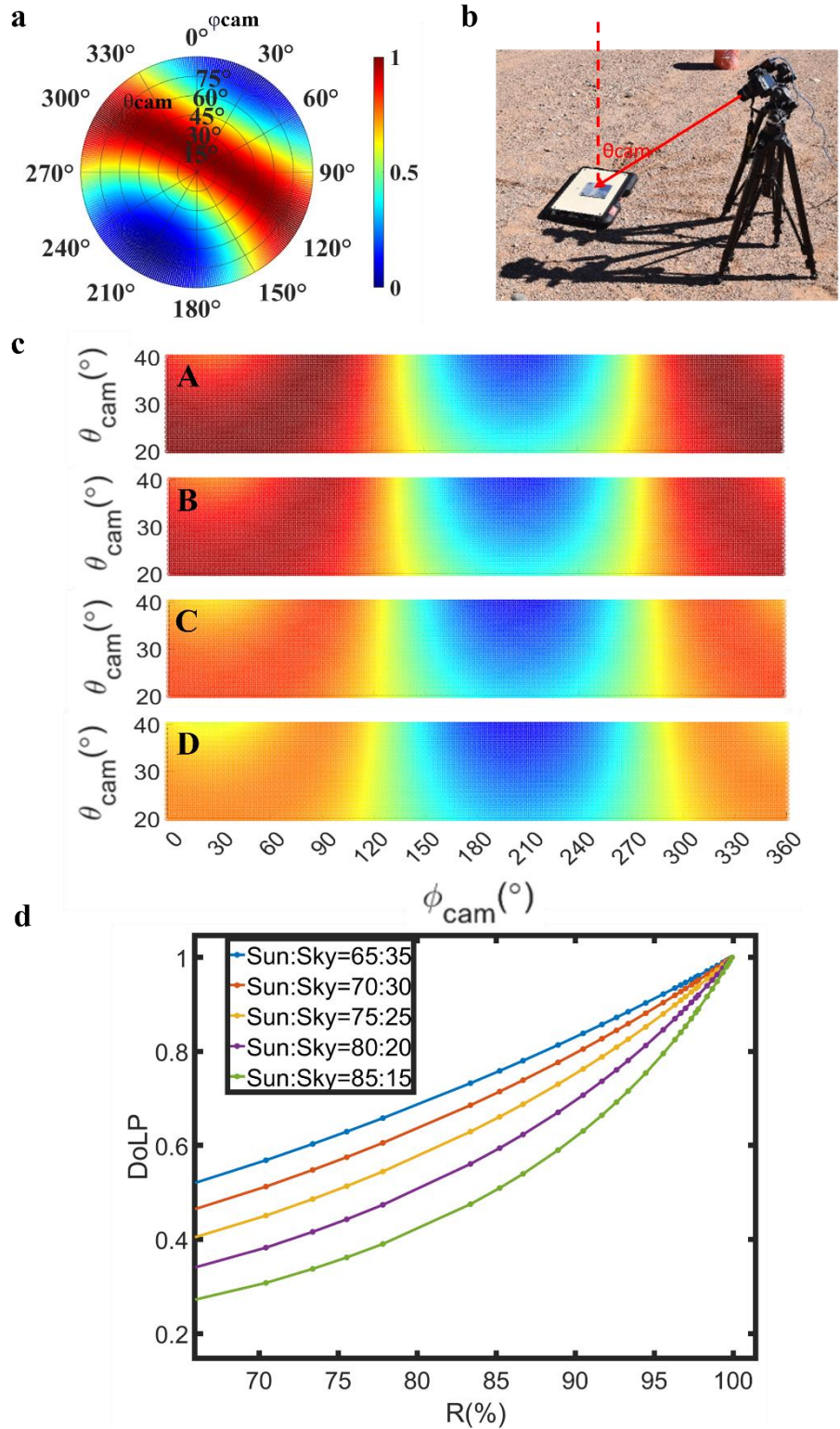

**Fig.5 Soiled Mirror Sample Measurement. a,** DoLP simulation of skylight reflection from a mirror positioned horizontally, with its surface normal pointing vertically upward. $\theta_{cam}$ and $\phi_{cam}$ are the zenith and azimuthal angles of the camera. **b,** Experiment

setup with polarization camera fixed on a tripod. **c,** Simulations of DoLP change as $\phi_{cam}$ changes with different reflection. $\theta_{cam}$ are limited to a 20 to 40 degrees window to show the area of interest captured by the imaging sensor. A, B, C, D denote different soiling regions in the simulation. **d,** Different fitting curves for this type of soil while varying the sunlight-to-skylight ratio.

The camera zenith angle $\theta_{cam}$ is 30°. From this simulation, we can expect that at this zenith angle, the azimuthal angles from 0° to 90° and 300° to 360° will likely have a good contrast of DoLP for different soiling levels. In the measurement, we perform an azimuthal scan at 15° increments to identify the direction in which the mirror exhibits the highest DoLP value. This region is cleaned during sample preparation and treated as the relatively cleanest area on the surface. Even if no absolutely clean area remains, our method defines reflectance (R) in a relative sense—normalized to the cleanest available region, as guided by simulation results (Fig. 5) showing a decrease in DoLP with increasing soiling. In this measurement, the Sun is at $(\theta_{sun}, \phi_{sun})$ = (68, 208) and camera is at $(\theta_{cam}, \phi_{cam})$ = (30, 300) on the celestial dome. Fig.5d shows the simulation curves of R vs DoLP with the same type of soil, camera position and acceptance angle, yet with different sunlight-to-skylight ratio. This ratio needs to be found by fitting the datapoints from the reference sample. We do not have a way to directly measure the sunlight ratio $K_{sun}$ and skylight ratio $K_{sky}$ as defined in Equation (18), but the theory suggests that this ratio affects noticeably the R-vs-DOLP curve, as shown in Fig.4c. Thus, we can carry out a calibration process by taking the DOLP images of the reference sample (known reflectance in different regions obtained by reflectance measurement) and fitting the R-vs-DOLP curve to obtain $K_{sun}$ and $K_{sky}$. At the same location, $K_{sun}$ and $K_{sky}$ mainly change as the weather changes (e.g. more cloud coverage) and time (different sun position). Unless there is a visible fast change of clouds, we assume $K_{sun}$ and $K_{sky}$ remain the same for each 15 to 30 minutes measurement window because the change of Sun position is not significant within this time period.

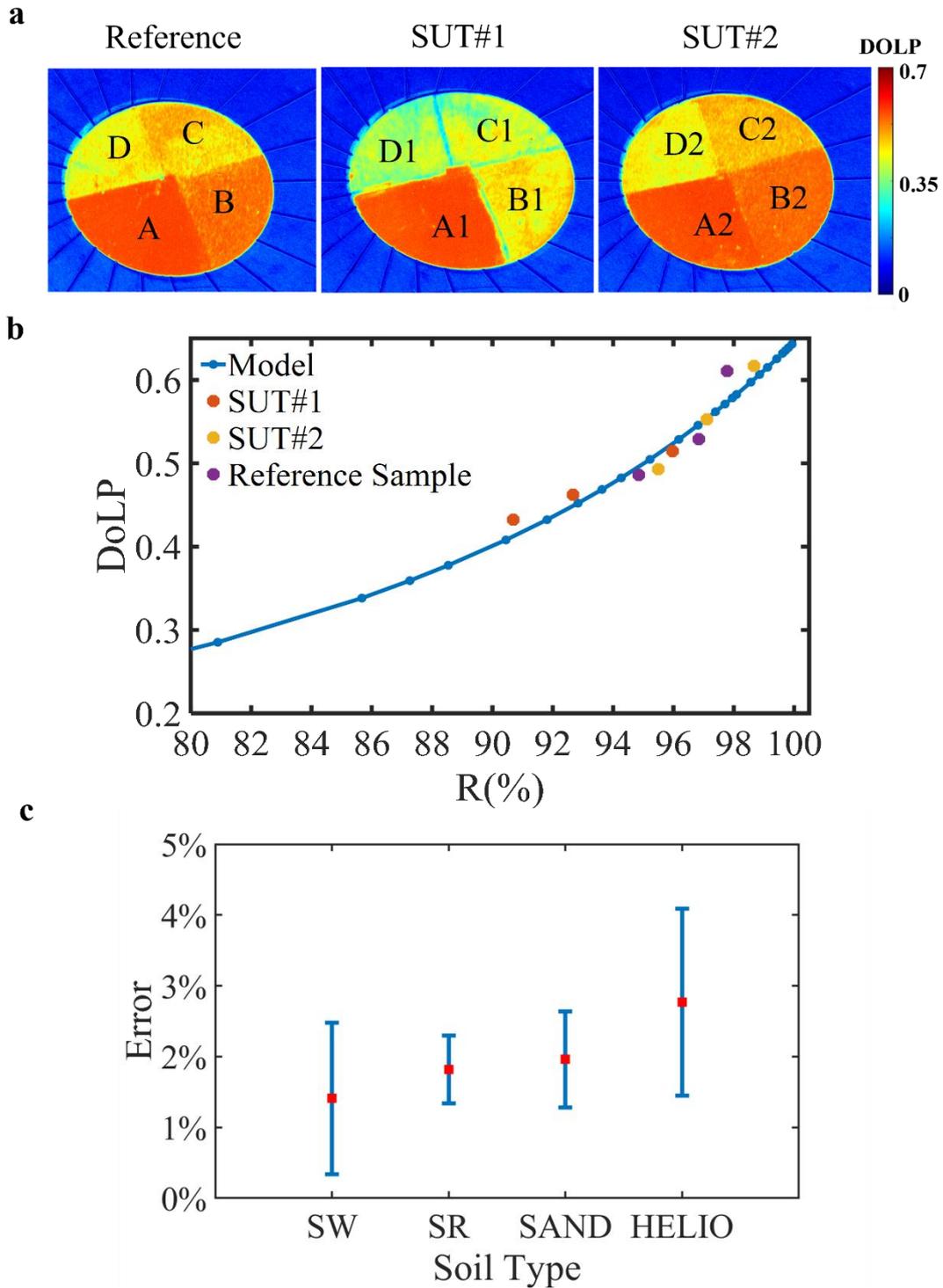

**Fig.6 Soiled Mirror Sample Measurement. a,** Captured DoLP images of the refence sample and the two Samples Under Test (SUT). **b,** Fitting curve of the four data points from the reference sample and the error of the two Sample Under Test (SUT). The sunlight-to-skylight ratio was acquired at 86% to 14%. Error was defined as the difference between the predicted reflectance from the fitted curve and the measured reflectance of the SUTs. Overall, the error was always under 3%. **c,** Measurement results for four different types of soil tested at ASU. The x-axis label corresponds to "Sahara White (SW)", "Sahara Red (SR)", "Sandia Soil (SAND)", "Sandia Heliostat Samples (HELIO)". Average errors of these four measurements were 1.41%, 1.82%, 1.96%, 2.77%.

The DoLP images for the reference sample and two Samples Under Test (SUT) are shown in Fig.6a. Notice that, in measurement, even the clean region that's designed to exhibit the highest DoLP region does not show DoLP value close to 1. This discrepancy from the theoretical model that the highest DoLP regions should have values close to 1 is due to several simplifications we made in the model. For the incident light, the simulation model only considers Rayleigh scattering forming skylight and there are no more scattering events as light travels in the air, reaches the mirror surface and gets reflected. Thus, in measurement, the highest DoLP values vary due to these neglected events. During the azimuthal scanning process, we select the image with highest DoLP value on the clean region in order to have the best contrast and scale all the sample images' maximum DoLP accordingly. The changing of DoLP at different azimuthal angles follows the trend in simulation (see Supplementary Fig.S4). Additionally, this also provides an insight into the flexibility of this method.

Thus, if the field test has constraints, such as a height limit for safety or a restricted operational area, the polarimetric imaging setup may not be able to access the region with the highest DoLP. In this case, similar to the weather conditions, the highest DoLP value captured will decrease yet can be still scaled accordingly to meet the fitting model's requirement. Another important factor to consider in data processing is the incidence angle. Although the camera angles that captured the reflected skylight with the highest incidence DoLP values are desired, one can easily find it is not always guaranteed in tests due to various reasons, such as safety requirements of the field operation, GPS deviation, deviation in heliostats tracking position, etc. However, the method developed is still valid to apply in the situations when camera is facing the region with lower incidence DoLP, by applying the Acceptance Angle Correction (AAC) described in Section 2.

As shown in Fig.6b, after the data was acquired, we analyzed the data by fitting the simulation curve's sunlight and skylight ratio parameters with the reference sample's DoLP and reflectance of the four regions. With the fitted curve, we input the DoLP values of each region on other samples and predict their reflectance. These Samples Under Test (SUT) were prepared in the same way as the reference sample with the same type of soil. Next, we compared the PIMS calculated reflectance and the measured reflectance to estimate the method's accuracy. In total, we have prepared and measured four different types of soil following the same procedures described above. Even though the fitting results are different for different types of soil and different dates of measurement, the error between the predicted reflectance using DoLP image and fitted model vs the measured reflectance was in the range of 1.41% to 2.77%, as shown in Fig.6c. The four types of samples measured were "Sahara

White (SW)", "Sahara Red (SR)", "Sandia Soil (SAND)", "Sandia Heliostat Samples (HELIO)". The first three were different types of soil deposited on to the same kind of silver mirror. The structure of the silver mirror is modeled as in Fig.2b. For the last type of sample, Sandia Heliostat Samples (HELIO), we do not have the accurate layer and material information. It was thus simulated similarly to the silver mirror samples, with a glass top layer, a reflective silver layer, and adding a copper layer at the bottom. It is possible that the larger error and standard deviation of this result come from the inaccuracy of the modeling in simulation.

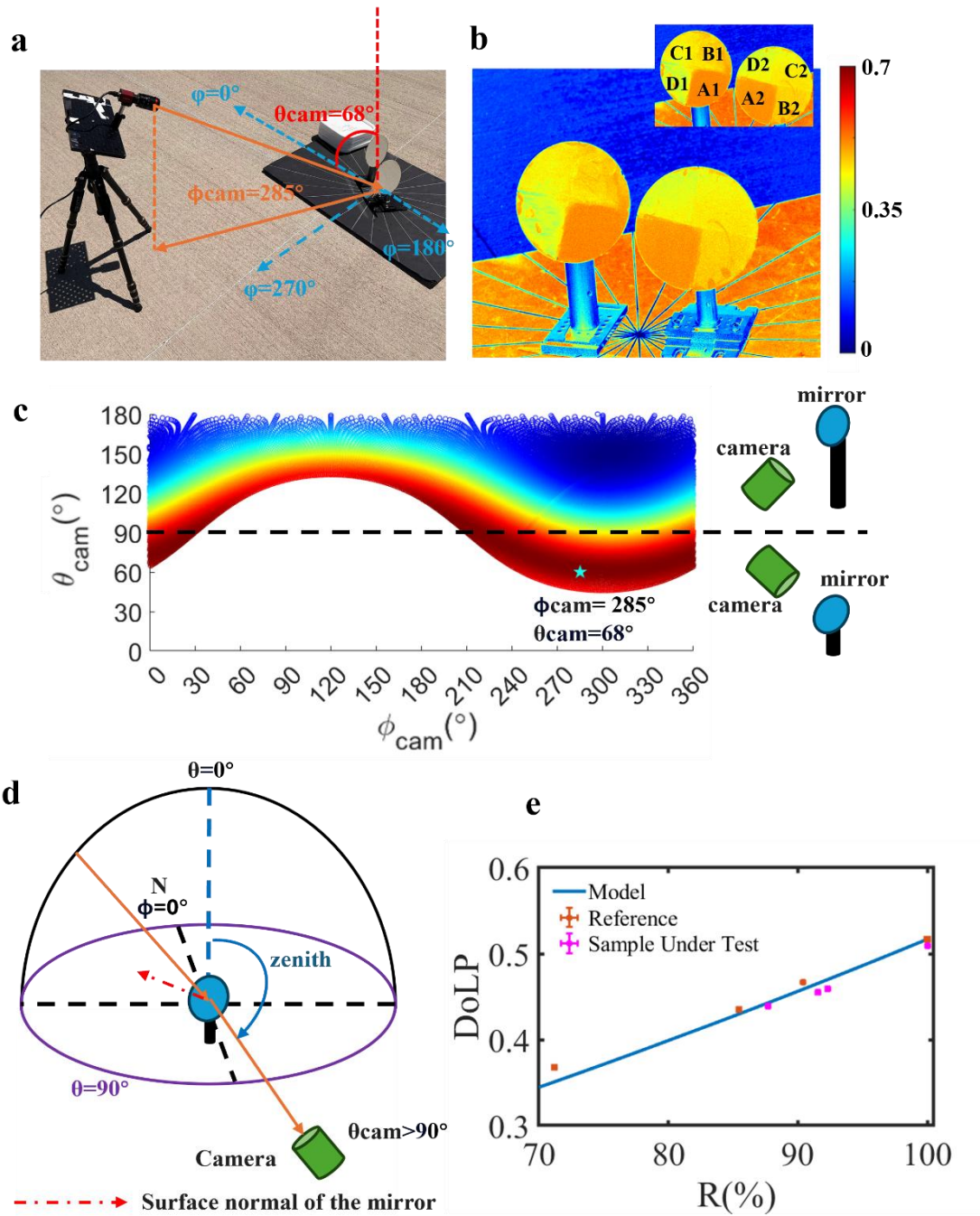

**Fig.7 Tilted Mirror Sample Measurement. a,** Measurement setup. Two soiled mirrors with the same type of soil were positioned at the same orientation. The zero point of azimuth angle starts at true North. The camera was placed at $\theta_{cam} = 68°$ and $\phi_{cam} = 285°$. **b,** Captured DoLP image. The regions of different soiling levels on each mirror can be distinguished clearly. **c,** Simulation of DoLP reflection from the mirror surface. Notice that since the mirrors have slanting angles, in the field it is possible to use zenith

angles larger than 90 degrees, meaning that camera is looking up, as indicated in the top half of the diagram. **d**, diagram showing when the camera is lower than the measured mirror, $\theta_{cam} > 90°$. **e,** fitted results using reference sample (left one in **b,**) and comparison with the sample under test (right one in **b**).

## *4.2. Measurements on Tilted Mirror Samples*

For testing in the CSP field, we expect the heliostat mirrors will be positioned at slanted angles corresponding with tracking positions. Thus, we designed experiments as shown in Fig.7a to further validate our model when the mirrors are tilted. From simulation of the reflected skylight off the tilted mirrors, we held the polarization camera at 68° zenith angle and did an azimuth scan to find the image with the highest DoLP values. Based on simulations of skylight reflection from tilted mirrors, we positioned the polarization camera at a zenith angle of 68° and performed an azimuthal scan to identify the viewing angle yielding the highest DoLP values. Fig. 7b shows the processed DoLP image corresponding to $\theta_{cam} = 68°$ and $\phi_{cam} = 285°$. During the measurement, the Sun position was $\theta_{sun} = 18.14°$ and $\phi_{sun} = 121.54°$, and the mirror surface normal was $\theta_{normal} = 25.60°$ and $\phi_{normal} = 276.98°$, as facing away from the Sun gets highest DoLP region of skylight. All angular measurements are expressed in the coordinate system defined with the center of the black cardboard (marked by white lines in Fig. 7a) as the origin, and with the +z axis oriented vertically upward (i.e., zenith direction). Both samples have the same type of soil, Sandia Soil, deposited as four regions with different soiling levels. The four regions are visually distinguishable from direct inspection. Since the mirror is tilted, the possible zenith range of its reflection towards camera extends beyond 90° as in the top half of Fig.7c. In the field, non-UAV alternatives such as handheld cameras or ground vehicles can still take advantage of this method while the camera is lower than the heliostat mirrors in altitude. In Fig.7d, we can calculate the camera position the same way as in Equation (12) but resulting in a $\theta_{cam}$>90°. This provides more flexibility of the method. While operating in the field, it is possible to have a handheld polarimetric imaging setup looking up to the area of interest. Fig.7e shows that the fitting results in a sunlight-to-skylight ratio of 65% to 35%. The relative reflectance fitting errors for the four regions of the reference sample are -0.072%, 1.249%, 0.898%, and 3.635% for region A1, B1, C1, D1, respectively. The relative reflectance fitting errors for four regions of the sample under test are -1.251%, -1.931%, -1.902%, and -0.675% for region A2, B2, C2, D2, respectively. Overall, the relative reflectance error is under 4%.

## 5. Field Tests

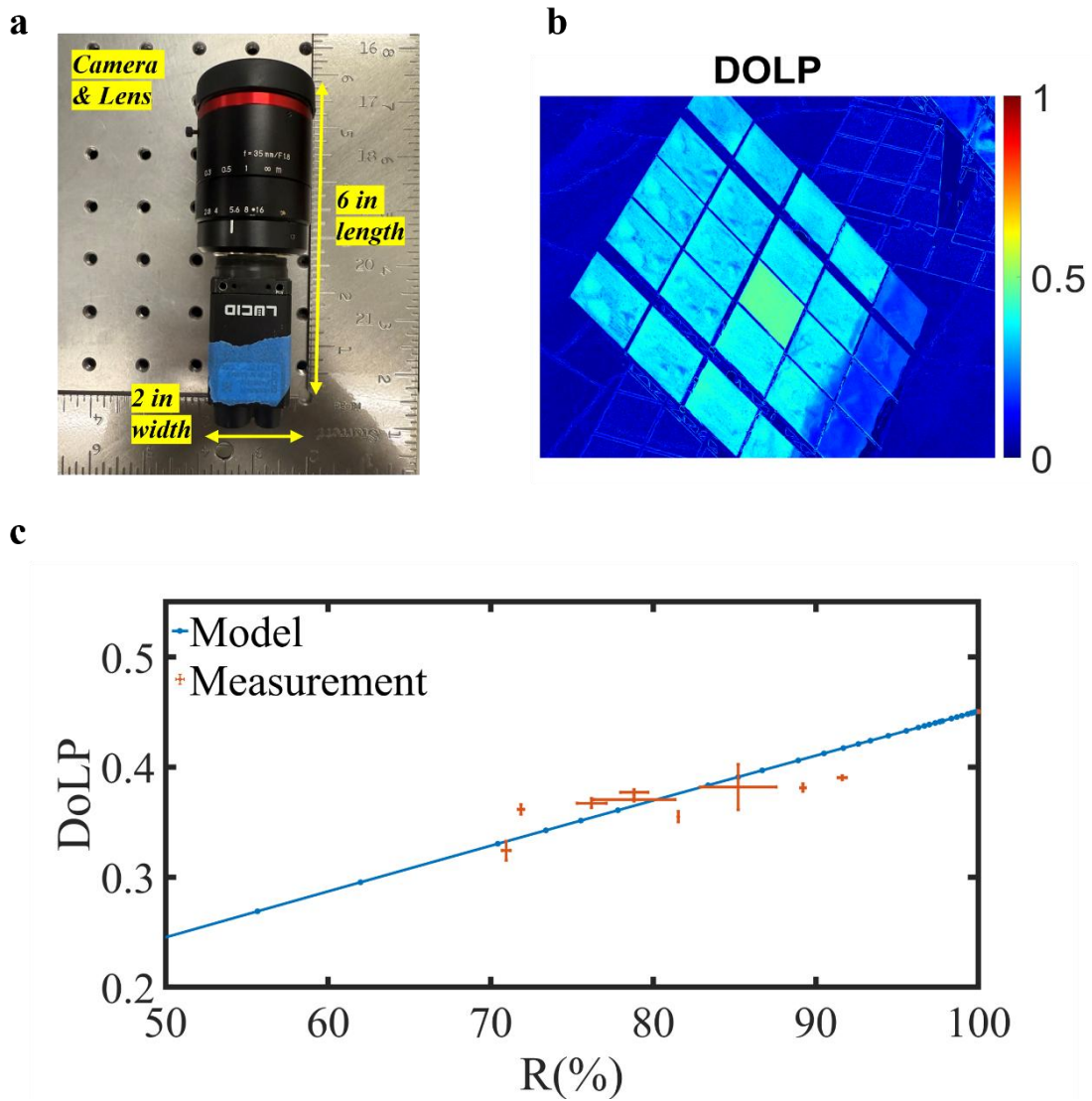

**Fig.8 Field test results on heliostat 14E2. a,** dimension of the polarization camera and the lens. **b,** DoLP image of heliostat 14E2. **c,** measurement results and the fitted DoLP-R curve with the data from 14E2.

So far, we have developed and validated the polarization-based reflectance prediction method. When it comes to field deployment, factors such as heliostat tracking, safety concerns, flight path or measurement positions and deployable systems need to be considered thoroughly. For CSP fields that require quick, large area soiling detection,

*38*

we designed an integrated polarimetric imaging drone system to execute precalculated waypoints and capture images of multiple target heliostats in one flight. This system is upgraded from our previous design [17]. The polarization camera (Lucid Triton) with a 35 mm lens (KOWA LM35HC 35 mm f/1.8) with 14.0° × 10.6° angle of view is integrated with a Jetson Xavier microprocessor to execute tasks such as live view, image capture and data storage. As shown in Fig.8a, this camera is portable and can be easily integrated into different systems according to the need. The system can be operated remotely with 15 fps video streaming and can take images by command. On the other hand, we measure the reflection of different facets with a calibrated Surface Optics 410-Solar Reflectometer on a boom lift (see Supplementary Fig.S5b). This measurement was done after the flight test and is designed to measure multiple points on each target facet to get reflectance information. We carried out field test at Sandia National Solar Thermal Test Facility (NSTTF) in Albuquerque, New Mexico using this setup. First, we captured images of heliostats with different soiling levels with the UAV-based setup. These images were captured while the test heliostats were at their Beam Characterization System (BCS) standby tracking position [27] to set the conditions such that a non-interruptive detection was carried out without interrupting the operation of the field. Since the tracking orientation and sun position at different times can be pre-calculated, simulation was done beforehand to find the designed position and direction of the camera for all times, sampled at 15 minutes intervals. Then for a given actual flight time, we selected the corresponding pre-calculated flight plan and flew the polarimetric imaging drone to the pre-calculated waypoints and adjusted camera orientations to capture polarization images with high DoLP values. The heliostats had their center facet cleaned up prior to the field test to provide the baseline of relative reflectance.

After obtaining and processing the data, we selected the part of the DoLP image for model fitting to be consistent with the region that was measured with Surface Optics 410-Solar Reflectometer. Then, by applying the simulation model, we calculated the fitted curve of R to DoLP. Because different mirror facets of the heliostats correspond to a different region of reflection, we adopted AAC for the DoLP and R values in the model fitting. Figure 8b shows the DoLP image of heliostat 14E2 (row 14, Column 2 East side of the tower). Figure 8c shows the fitting curve ("Model") and the measurement results ("Measurement"). During this measurement, the center of the heliostat 14E2 was set as the origin, with Sun position at $\theta_{sun} = 22.75°$ and $\phi_{sun} = 239.28°$. UAV camera's position is $\theta_{cam} = 68.31°$ and $\phi_{cam} = 118.90°$. At the Sandia NSTTF, heliostats in BCS standby mode are tracking a point located at (x = 60 m, y = 8.8 m, z = 28.9 m) in the site's coordinate system. The coordinate origin is defined at the base center of the tower, with the z-axis aligned with the zenith (upward direction), and the x- and y-axes pointing east and north, respectively. The data points are acquired by reflectometer measurement (R) and DoLP measurement (average DoLP in the region of interest). For each datapoint, the horizontal error bar represents the standard deviation of repeated reflectance measurements using the Surface Optics 410-Solar reflectometer, and the vertical error bar indicates the DoLP standard deviation calculated from the selected region in the DoLP image. As described in Section 2.2, any reflectometer datapoints with a standard deviation exceeding 5% were excluded from analysis due to measurement instability. We performed the same reflectometer measurements on smaller mirror samples in the lab, where all data points showed standard deviations below 5%. In contrast, field measurements conducted from a boom lift exhibited greater variability, with some datapoints showing standard deviations greater than 5% and thus excluded from further analysis. This variation is likely due to a combination of factors: mechanical instability while manually holding the reflectometer against large heliostat facets (see Supplementary Fig. S5b), and inhomogeneity in natural soiling, which can lead to measurable reflectance differences if the measurement spot shifts slightly between repetitions. These challenges highlight the limitations of manual reflectometer use under real-world field conditions. The captured DoLP results, and the measured reflectance values from reflectometer measurement were then used to fit the model and calculate the error. The center facet was cleaned before the measurement. The low DoLP values observed in the last row of mirrors in Fig. 8b results from their reflection of the surrounding mountains rather than the sky. This is not indicative of a positional error; instead, it illustrates a common scenario where unintended objects appear in reflection images, leading to measurement discrepancies. Such limitations naturally arise when employing UAV-based methods while heliostats are actively tracking. In practical applications, we can include the surrounding objects as part of the flight path calculations to avoid this.

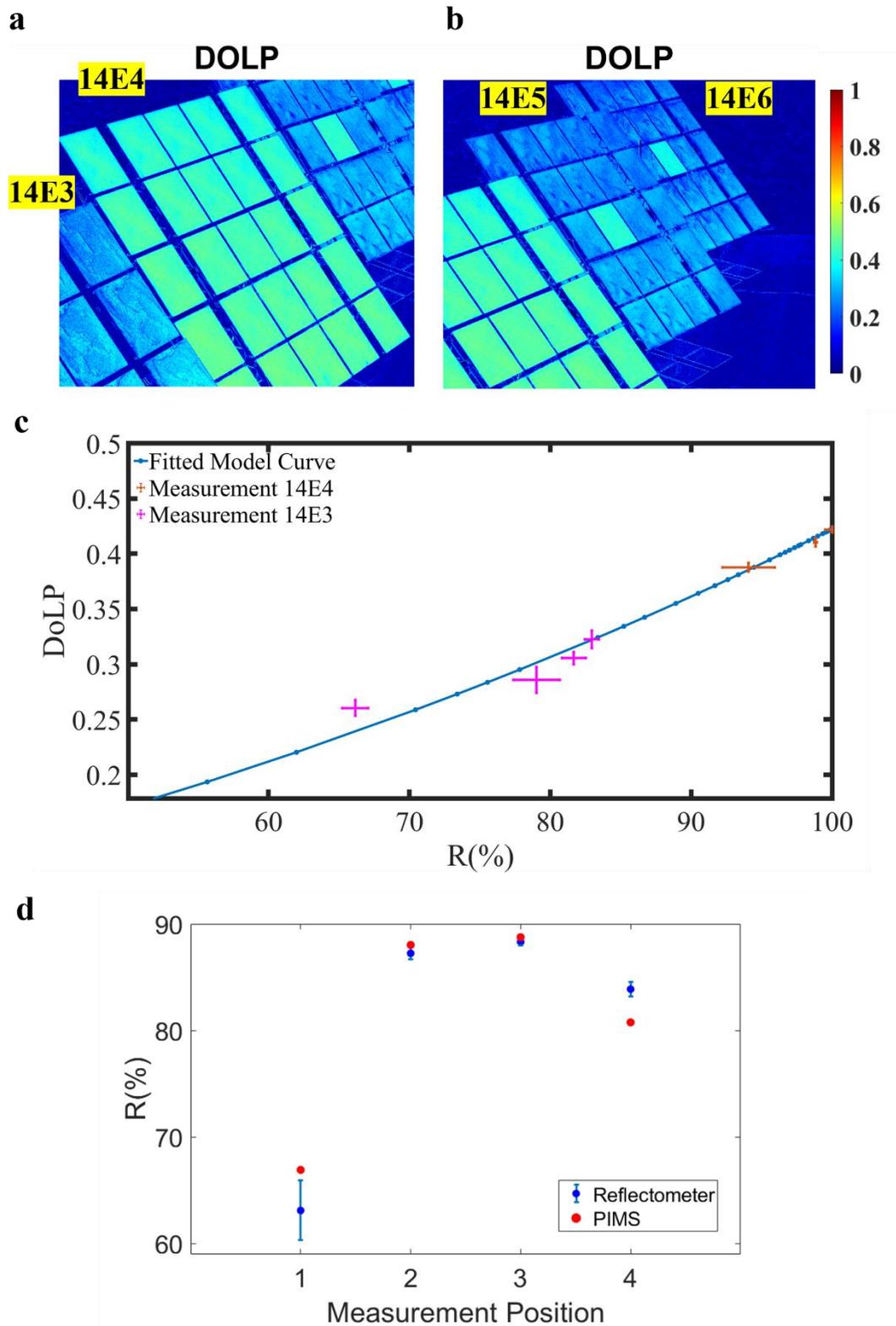

**Fig.9 Field test results on heliostat 14E3, 14E4, 14E5 and 14E6. a,** DoLP image of heliostats 14E3 and 14E4 used as clean reference mirrors. **b,** DoLP image of heliostats 14E5 and 14E6 used for performance testing. **c,** Relationship between DoLP and relative reflectance derived from the reference mirrors. This fitting curve is applied to estimate reflectance from the DoLP values of 14E6. **d,** Comparison between reflectance values derived from PIMS (based on DoLP) and ground-truth measurements from Surface Optics 410-Solar Reflectometer for heliostats 14E5 and 14E6. The spatial positions of the sampled PIMS points are shown in Supplementary Fig. S6.

In another flight, we captured and processed DoLP image of heliostat 14E3 and 14E4 (Fig.9a) and DoLP image of heliostat 14E6 (Fig.9b). The position of each measured data point is indicated in Supplementary Fig.S6. We used the DoLP data points of 14E3 and 14E4 to form a reference set. During this measurement, the center of the heliostat 14E4 was set as the origin, with Sun position at $\theta_{sun} = 42.41°$ and $\phi_{sun} = 264.16°$. UAV camera's position is $\theta_{cam} = 62.14°$ and $\phi_{cam} = 102.67°$. The heliostats are tracking BSC standby point. Their corresponding relative reflectance (R) data was measured using Surface Optics 410-Solar Reflectometer. Similar to the 14E2 results, for each datapoint, the horizontal error bar indicates the R standard deviation, and the vertical error bar indicates the DoLP standard deviation. The fitted curve, with $K_{sky} = 40\%$ and $K_{sun} = 60\%$, has a fitting Mean Absolute Error (MAE) for R of 2.04% with standard deviation of 2.70% The error here is defined as the difference between the curve's R at the same level of DoLP acquired polarization image and the R measured by reflectometer. Then, we apply the acquired DoLP values from 14E6 (Fig.9b) to the fitted curve and get the results of relative reflectance calculated using PIMS. We show the comparison with the measured R from reflectometer in Fig.9d. The four data points on 14E6 now have a MAE of 2.04% with a standard deviation of 1.45%, where the error is defined as the difference between PIMS calculated R and reflectometer measured R. Generally, we evaluate the accuracy of PIMS on heliostat measurement shows the error of R less than 3%, making it useful for soiling level detection in CSP.

## 6. Conclusion and Outlook

We have demonstrated a field deployment method for high-speed, non-interruptive measurement of the heliostat mirror soiling status based on polarimetric imaging, i.e., the Polarimetric Imaging-based Mirror Soiling (PIMS) detection method, including an imaging setup, an experimentally calibrated model and measurement procedures. The imaging setup is portable and can be integrated onto a UAV. The measurement procedures are quick and simple without the requirement of additional light sources other than sunlight and skylight. The experimentally calibrated PIMS method enables the mapping of mirror reflectance with high spatial resolution, comparable to that of conventional imaging sensors. PIMS achieves a reflectance error below 3%. We have carried out mirror soling measurements in outdoor environments with different types of soils and shown that our method is applicable to different solar fields with calibration and fine-tuning model parameters such as soiled particle size distribution, refractive indices of the mirror surface and soiled particles, etc. Preliminary field tests carried out at Sandia NSTTF demonstrated that the polarimetric imaging drone-based mirror soiling detection can scan multiple heliostat mirror facets in single flight. Compared to state-of-the-art methods, the PIMS is rapid, non-intrusive to field operation, labor- and cost-efficient as it requires minimum equipment installation and operators.

The PIMS also has portability in different field conditions. In our measurements, we have a pre-cleaned region. However, an ideal design will be that we have a reference sample, same as the samples prepared in Section 3.1. It should have a clean region and at least three regions with different soiling levels. We can pre-measure this sample as reference and then start the drone flight. On the other hand, we can also pre-install smaller reference mirror facets close to the heliostats and clean them before measurement. During the drone flight, we will also capture the reference mirror facets image to be used as "clean region". For solar fields that require detailed soiling detection on individual mirror facets, or limit the use of drones, a portable setup that can be manually carried or installed on a ground vehicle to capture the polarization images is applicable. This method holds the potential to enable the autonomous monitoring of the CSP field soiling status over a large area, with fast scanning, accurate prediction, and low labor requirement. Furthermore, the simulation model and measurement procedures of PIMS can also apply to other reflective surfaces other than heliostat mirrors, and different particles other than soil. It can be further developed to accommodate other types of solar fields, such as parabolic troughs.

It is important to note that the preliminary field tests lack reference sample measurements for validation due to limited resources and field time available so far. An optimized field test workflow is illustrated in Fig.10a. We measure a reference sample independently from the UAV flight measurement. R and DoLP are measured using Surface Optics 410-Solar Reflectometer and a polarization camera, respectively. Due to the potential change of skylight and sunlight ratio, we redo the measurement each 30 minutes. Within the 30 minutes window, the skylight DoLP pattern does not change significantly, and thus the same ratio can be applied. The DoLP images taken with the UAV flight are then used to calculate R for different heliostats in the images, based on the curve fitted with the reference sample datapoints.

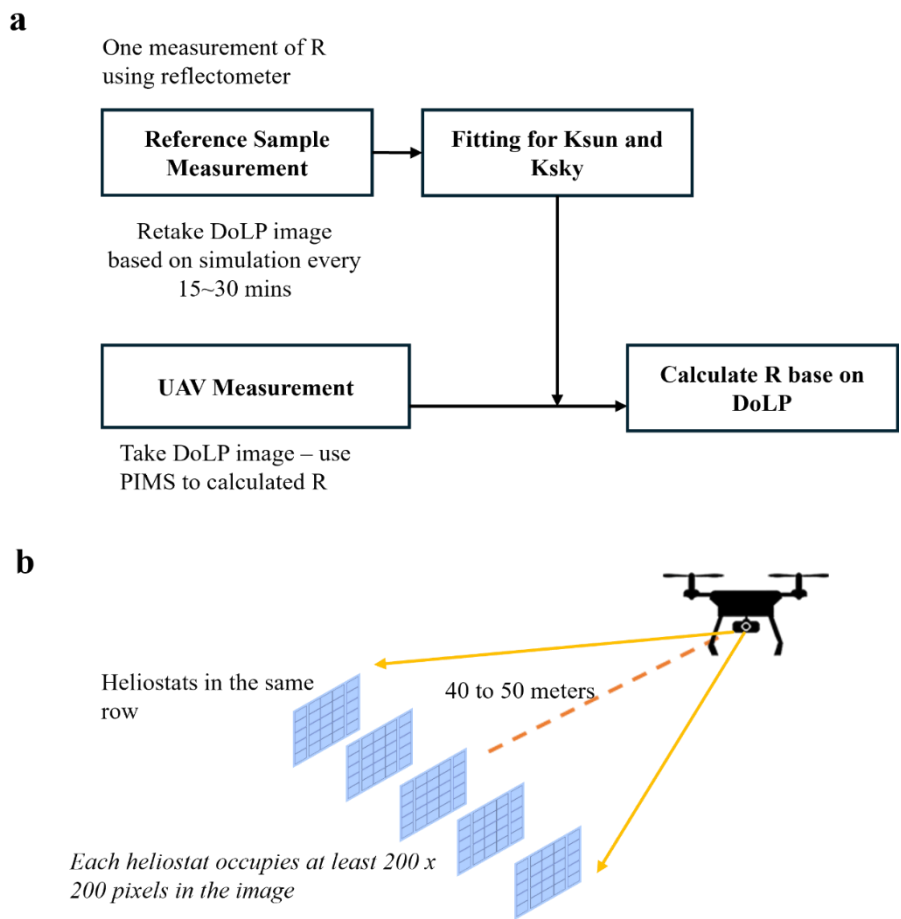

**Fig.10 Optimized field test flow chart the estimated capacity. a,** an optimized flowchart describing the measurement procedures. **b,** an estimation of measruement

capacity using UAV-based PIMS. Placing the drone from 40 to 50 meters away, we can measure at least 5 heliostats in each image.

The UAV-based setup used for the field tests presented in Section 5 is a prototype designed to demonstrate the feasibility of the PIMS method. It offers only limited hardware functionality. While the operator is able to view a live feed from the onboard camera via a wireless connection, the polarization images must be processed post-flight due to transmission bandwidth constraints. These current limitations are not inherent to the PIMS method itself. We estimate that with a mature design for commercial use, it is possible for the UAV setup to capture larger area with multiple heliostats in each image and thus evaluate the soiling level of different heliostats across the field more efficiently, as shown in Fig.10b. For the heliostats in Sandia NSTTF, we can estimate that 5 to 6 heliostats can be measured in each image and a ~30 mins flight can measure up to 100 heliostats, theoretically.

## 7. Acknowledgements

The authors gratefully acknowledge Neel Desai, Vishnu Pisharam, Dongyao Wang, and Zengyu Cen from Arizona State University (ASU) for their invaluable assistance with data collection during outdoor measurements, Dr. Fang Li from ASU for her expertise and help in the preparation of soiled mirror samples, Maddie Hwang from Sandia National Labs for her contribution to soil sample collection at the Sandia National Solar Thermal Test Facility (NSTTF), Luis Garcia Maldonado and Zach Bernius from Sandia National Labs for their support during field tests at the Sandia NSTTF.

## 8. Data Availability

Data available on request from the authors. MATLAB code for data processing is available at https://github.com/motian-asu/PIMS.

## 9. CRediT authorship contribution statement

**Mo Tian**: Data curation, Formal analysis, Methodology, Field test, Software, Visualization, Writing – original draft, Writing – review & editing. **Md Zubair Ebne Rafique**: Conceptualization, Formal analysis, Methodology, Software. **Kolappan Chidambaranathan**: Methodology, Software. **Randy Brost**: Project administration, Resources, Field test, Writing – review & editing. **Daniel Small**: Resources, Software, Field test. **David Novick**: Resources, Software, Field test. **Julius Yellowhair**: Methodology. **Yu Yao**: Conceptualization, Formal analysis, Funding acquisition, Writing – review & editing, Resources, Supervision, Project administration.

## 10. Funding

The research conducted at Arizona State University is supported by the U.S. Department of Energy Solar Energy Technology Office under the contract no. EE0008999. Sandia National Laboratories is a multi-mission laboratory managed and operated by National Technology & Engineering Solutions of Sandia, LLC, a wholly-owned subsidiary of Honeywell International Inc., for the U.S. Department of Energy's National Nuclear Security Administration under contract DE-NA0003525.

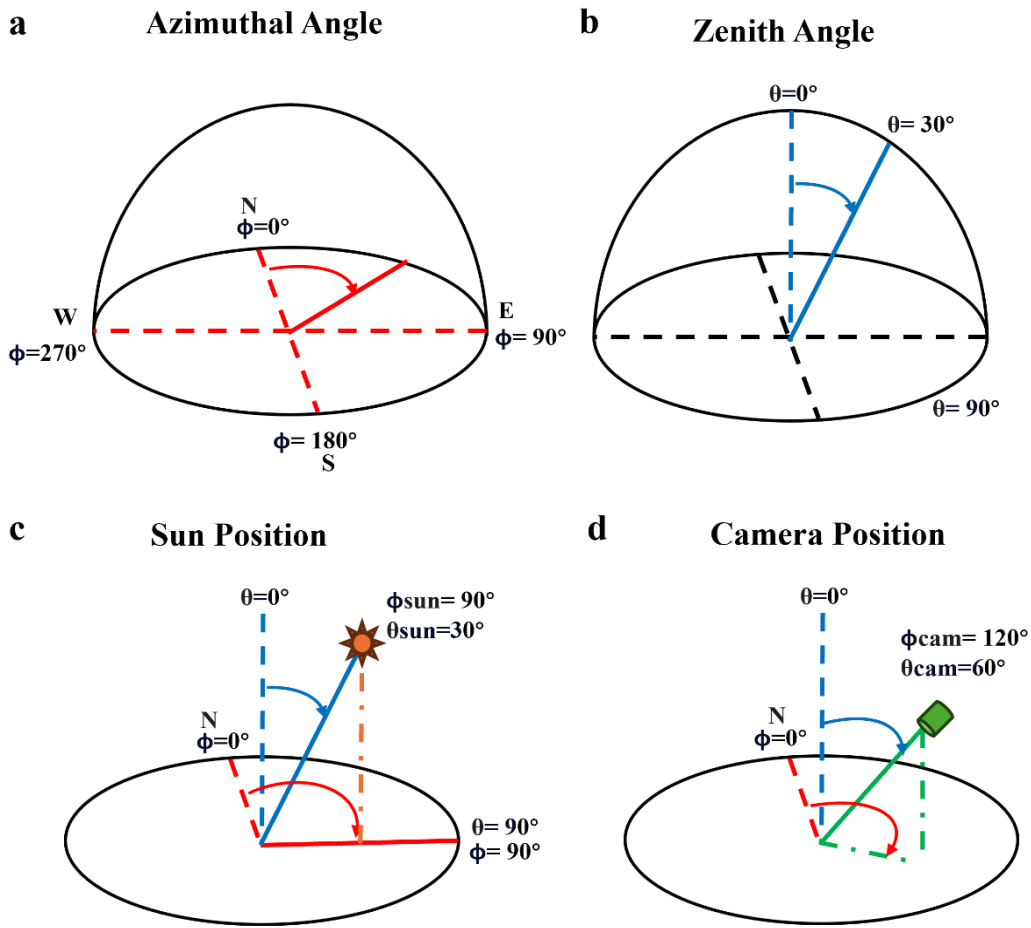

**Supplementary Fig S1. Definition of the global coordinate system. a,** azimuthal angle starts from North and increases clockwise. **b,** zenith angle. **c,** sun position definition in the global coordinate system. **d**, camera position definition in the global coordinate system.

*51*

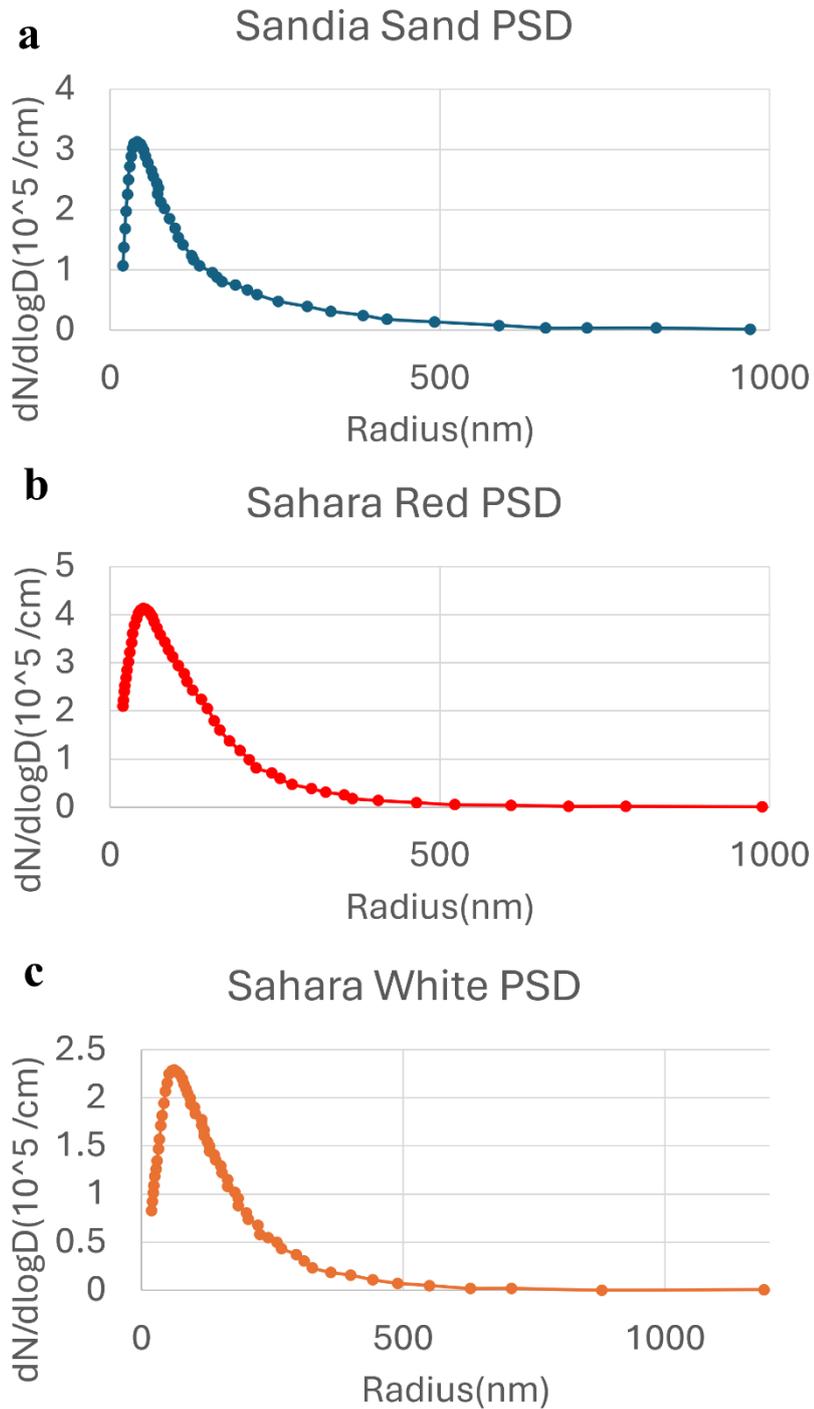

**Supplementary Fig.S2. Particle Size Distribution (PSD) of the three types of soil used in the measurement. a,** soil sample from Sandia NSTTF. **b,** soil sample from Sahara Desert with red color. **c,** soil sample from Sahara Desert with white color.

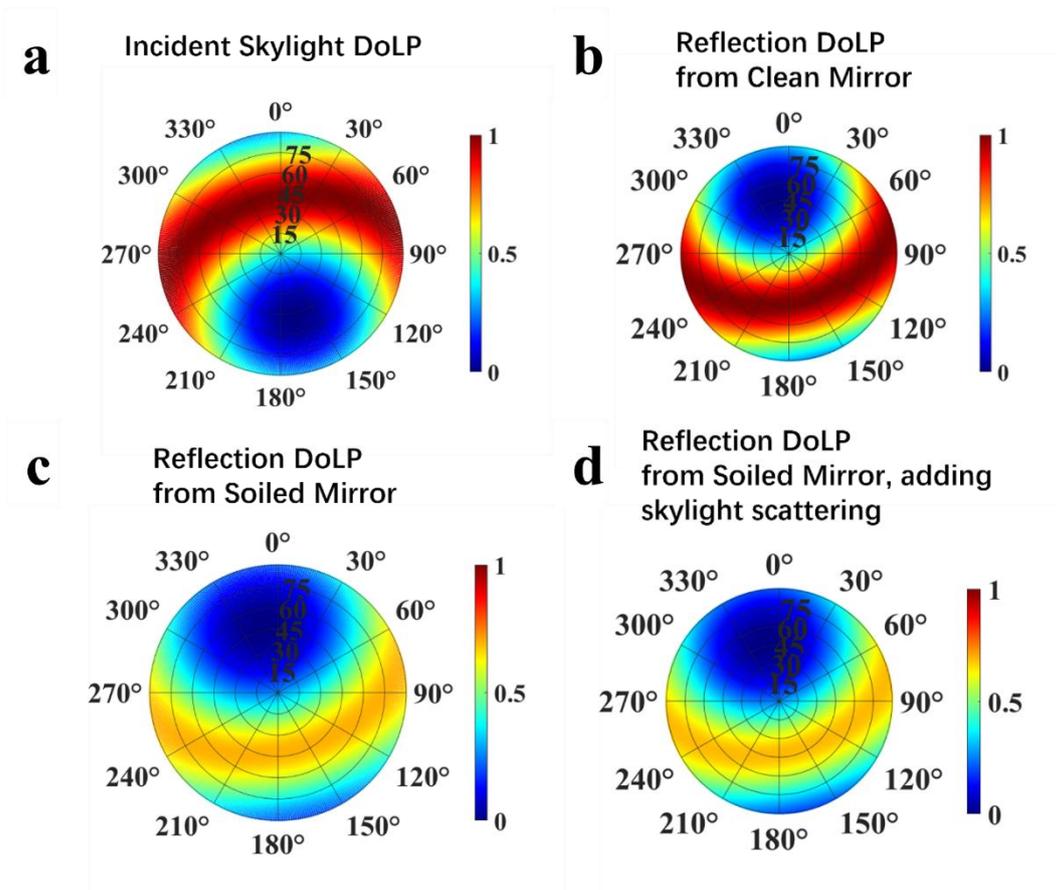

**Supplementary Fig.S3 Simulation results comparison if adding the neglected terms in the sunlight-skylight model. a,** simulation of the incident skylight. **b,** simulation of the reflected light with the mirror surface at 100% reflectance (i.e. clean mirror). **c,** simulation of the reflected light with the mirror surface at 90% reflectance, only considering skylight reflection from the clean region and sunlight scattering from the soiled region, as described in the model. **d,** simulation of the reflected light with the mirror surface at 90% reflectance, considering skylight reflection from the clean region, skylight scattering and sunlight scattering from the soiled region.

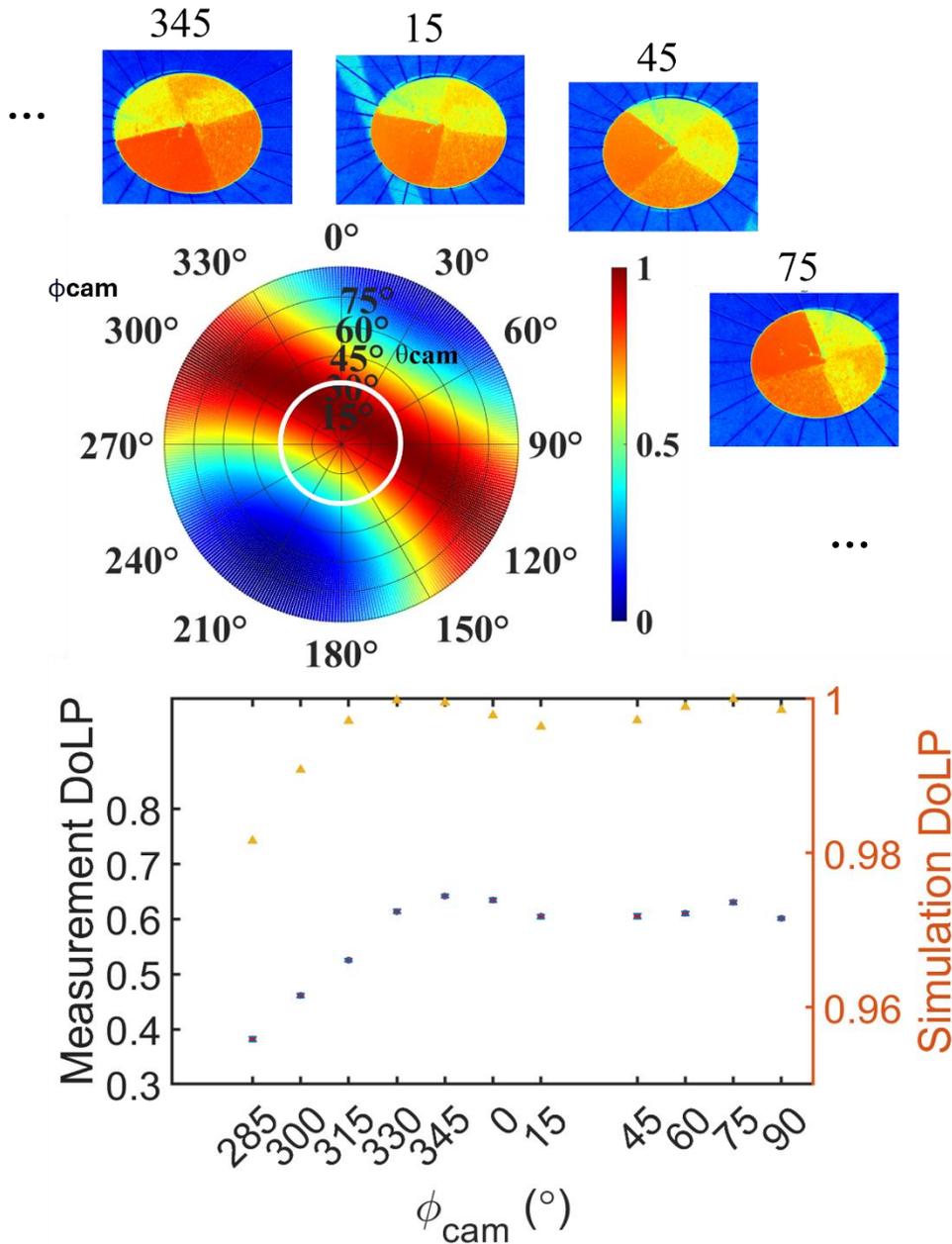

**Supplementary Fig.S4 DoLP comparison between simulation and measurement for leveled mirror sample at same zenith but different azimuth.** Measurements were taken as a camera azimuthal scan with same camera zenith angle at $\theta_{cam}$ =30°. At different ϕcam, simulation can predict the change of DoLP due to different incidence, and thus even we did not get the image with the highest DoLP, PIMS method can still be applied to the image to calculate R.

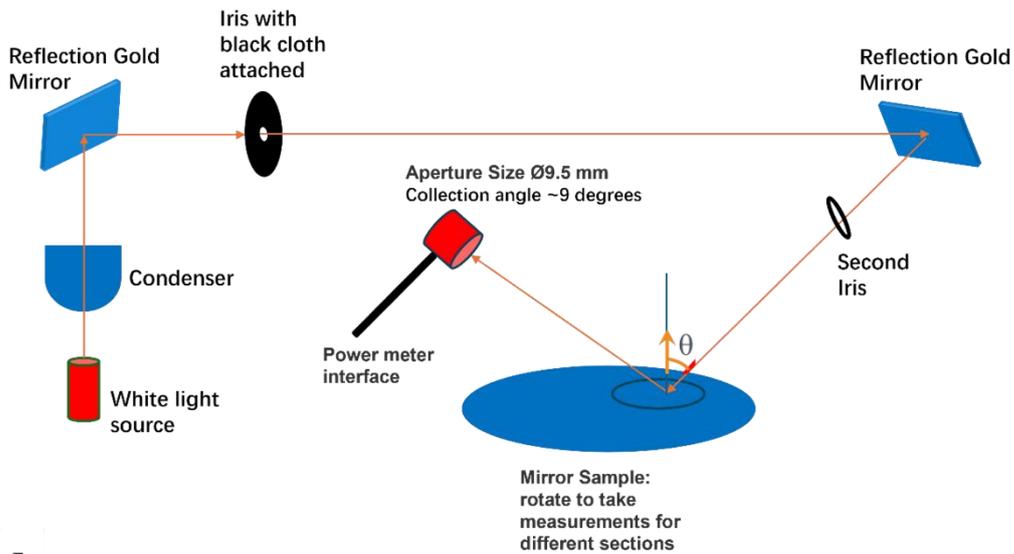

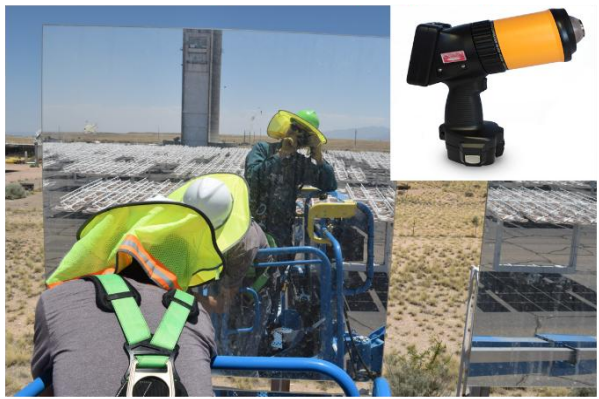

**Supplementary Fig.S5 Reflection measurement setups. a,** reflection measurement setup diagram for small soiled mirror samples in lab. For each measurement, the incident angle is taken the same as the azimuth angle in the outdoor measurement for capturing polarization images. The clean region's power meter reading is taken as the baseline. As we rotate to different regions of the mirror with different soiling levels, the power meter reading varies, and the ratio of soiled region readings versus clean region readings is taken as the relative reflectance data. **b,** in-field measurement using reflectometer on a boom lift.

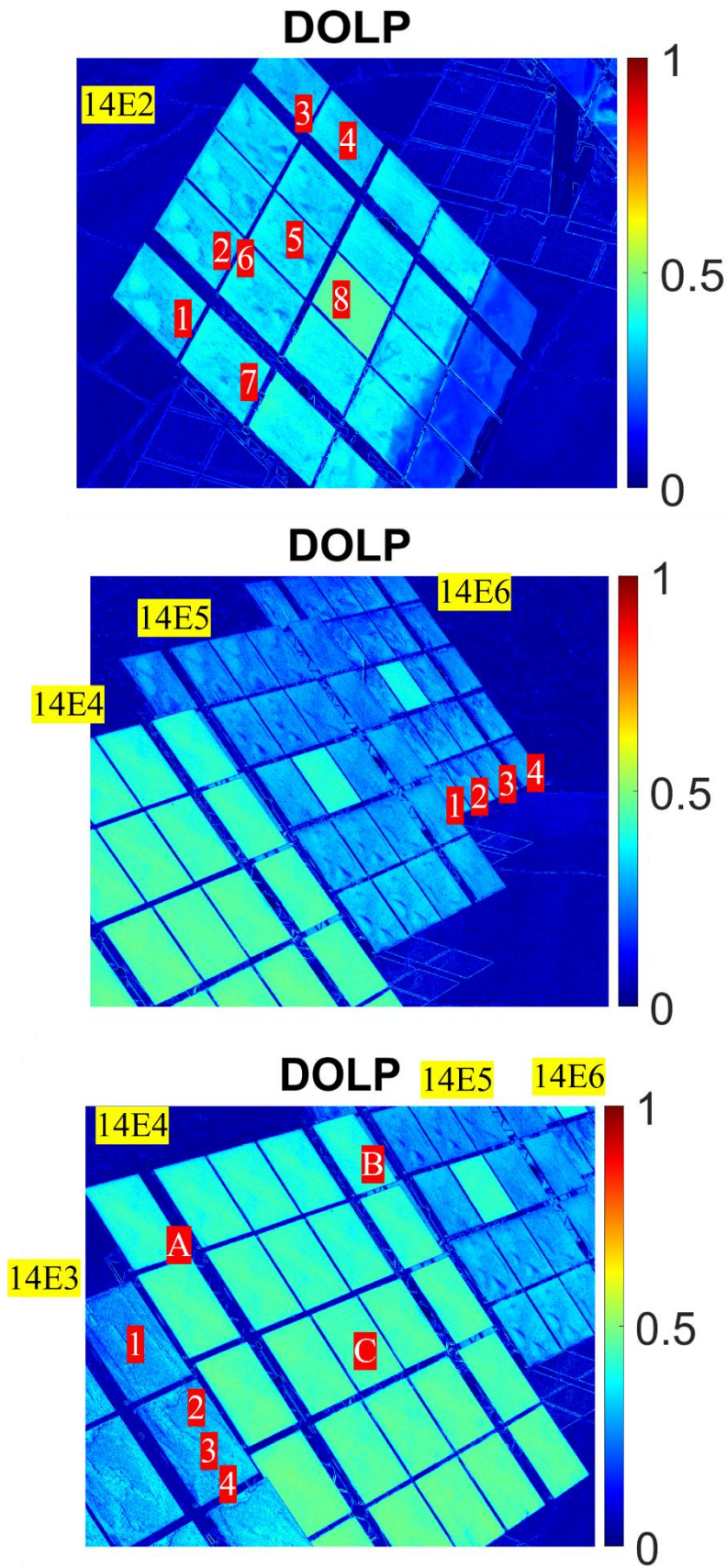

**Supplementary Fig.S6** Measured datapoints location for the field test images.